\documentclass[pre,reprint,final,floatfix,nofootinbib]{revtex4-2}

\makeatletter

\usepackage[T1]{fontenc}
\usepackage{mathtools}

\usepackage{microtype}
\usepackage{color}
\definecolor{bluesmoke}{rgb}{0.2,0.4,0.6}
\usepackage[
  allcolors=bluesmoke,
  colorlinks,
  pdftex,
  pdftitle={Geometric Localization of Waves on Thin Elastic Structures},
  pdfauthor={Manu Mannattil and Christian D. Santangelo},
  pdfkeywords={elastic waves, localization, WKB/semiclassical methods},
  pdfsubject={wave motion, mechanics, applied mathematics},
  hypertexnames=false
]{hyperref}
\usepackage{graphicx}
\graphicspath{{figures/}}

\usepackage[widespace]{fourier}
\DeclareMathAlphabet{\mathscr}{FMS}{futm}{m}{n}
\SetMathAlphabet{\mathscr}{bold}{FMS}{futm}{b}{n}
\DeclareMathAlphabet{\mathcal}{OMS}{lmsy}{m}{n}
\SetMathAlphabet{\mathcal}{bold}{OMS}{lmsy}{b}{n}
\usepackage[scale=0.9]{tgheros}
\DeclareMathAlphabet{\mathsf}{\encodingdefault}{\sfdefault}{\mddefault}{n}
\SetMathAlphabet{\mathsf}{bold}{\encodingdefault}{\sfdefault}{\bfdefault}{n}
\usepackage{bm}
\def\big{\bBigg@{1}}
\def\Big{\bBigg@{1.5}}
\def\bigg{\bBigg@{2.4}}
\def\Bigg{\bBigg@{3.2}}

\usepackage{addlines}

\allowdisplaybreaks

\def\bibsection{%
  \par
  \begingroup
  \baselineskip26\p@
  \bib@device{\hsize}{72\p@}%
  \endgroup
  \nobreak\@nobreaktrue
  \addvspace{19\p@}%
}%

\definecolor{hlcolor}{rgb}{0.68,0.1,0}

\DeclareMathOperator{\sech}{sech}
\DeclarePairedDelimiter{\abs}{\lvert}{\rvert}
\DeclarePairedDelimiter{\Abs}{\|}{\|}
\def\dd{\ensuremath\mathrm{d}}

\def\trans{\ensuremath{^\mathsf{T}}}
\DeclareMathOperator*{\argmin}{argmin}

\g@addto@macro\itemize{\setlength{\itemsep}{0pt}}
\g@addto@macro\enumerate{\setlength{\itemsep}{0pt}}
\g@addto@macro\description{\setlength{\itemsep}{0pt}}

\makeatother

\begin{document}

\title{Geometric Localization of Waves on Thin Elastic Structures}
\author{Manu Mannattil}
\email{mmannatt@syr.edu}
\author{Christian D.~Santangelo}
\affiliation{Department of Physics, Syracuse University, Syracuse, New York 13244, USA}

\begin{abstract}
We consider the localization of elastic waves in thin elastic structures with spatially varying curvature profiles, using a curved rod and a singly curved shell as concrete examples.  Previous studies on related problems have broadly focused on the localization of flexural waves on such structures.  Here, using the semiclassical WKB approximation for multicomponent waves, we show that in addition to flexural waves, extensional and shear waves also form localized, bound states around points where the absolute curvature of the structure has a minimum.  We also see excellent agreement between our numerical experiments and the semiclassical results, which hinges on the vanishing of two extra phases that arise in the semiclassical quantization rule.  Our findings open up novel ways to fine-tune the acoustic and vibrational properties of thin elastic structures, and raise the possibility of introducing new phenomena not easily captured by effective models of flexural waves alone.
\end{abstract}

\maketitle

\section{Introduction}
\label{sec:introduction}

Studying the propagation of elastic waves on thin structures is of crucial importance to a variety of problems in science and engineering, with applications ranging from acoustic cloaks to negative refraction~\cite{farhat2009,craster2012,zangeneh-nejad2019}.
Of particular relevance to many of these applications are localized waves, which are time-harmonic solutions to a wave equation that remain confined to a certain region of space without needing a confining potential or force.
Indeed, such waves are observed in many physical systems where they are often caused by heterogeneities in the medium or the boundary.
For instance, the Helmholtz equation admits bound states in arbitrary dimensions when solved on a tubular domain, provided that the tube is not everywhere straight~\cite{goldstone1992}.
Likewise, in waveguides in the form of an elastic plate, described again by coupled Helmholtz equations, waves localize around points of maximal curvature~\cite{gridin2005}.
Bound waves of similar nature have also been predicted in waveguides in the form of rods~\cite{gridin2005a}, elastic strips with varying elastic moduli~\cite{forster2006} and thickness~\cite{postnova2008}, quantum waveguides~\cite{duclos1995}, etc.

Localized waves can also arise in elastodynamic systems described by higher-order wave equations.
In this context, \citet{scott1992} studied the localized vibrations of a musical saw---an ordinary hand saw bent into the shape of the letter \textsf{S} and playable like a musical instrument~\cite{leonard1989,stuckenbruck2016}.
More recently, \citet{shankar2022} revisited the musical saw using both experiments and theory.
Forgoing an explicit analytical computation of the mode frequencies, they argued that the bound modes that appear at the inflection point of the saw are topologically protected.

Despite the efforts of the aforementioned authors, several critical aspects of wave localization on thin structures remain unclear:
What kinds of structural geometries support trapped waves, and of what types?
Are the vibrational spectra of one- and two-dimensional thin structures different?
Is it possible to compute, even approximately, the shapes and frequencies of the trapped modes?
In this paper, we explore these questions, choosing a singly curved shell and a curved rod as our examples.

\subsection{Background}
\label{sec:intro_background}

\begin{figure}
  \begin{center}
    \includegraphics{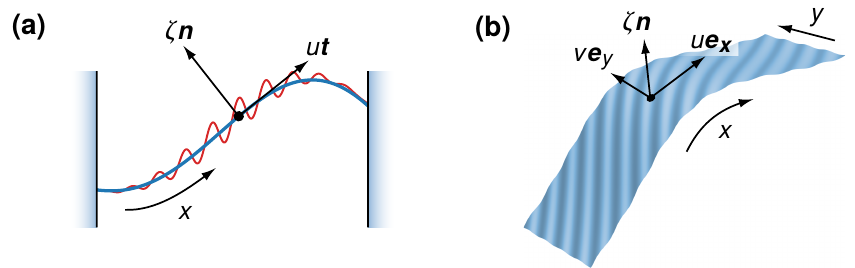}
 \end{center}
  \caption{Waves can propagate on thin elastic structures such as (a) rods and (b) shells.
    The undeformed structure is parameterized by the coordinates $x$ and $y$.
    Curvature couples tangential displacements ($u$ and $v$) that stretch/shear the structure with normal displacements ($\zeta$) that bend it.}\label{fig:waves}
\end{figure}

If a thin structure is uncurved, there are three basic types of waves that can propagate.
Extensional waves propagate by stretching and compressing the structure, and involve only the tangential displacements ($u$ and $v$ in Fig.~\ref{fig:waves}).
Flexural waves, by contrast, propagate by bending the structure and involve only the normal displacement ($\zeta$ in Fig.~\ref{fig:waves}).
In flat plates, shear waves, which do not compress or expand the plate, and involve only the tangential displacements propagate as well~\cite{landau1986}.

The situation gets complicated when the structure is curved.
First, curvature tends to couple the tangential and normal displacements, and therefore, we can only speak of waves that are predominantly flexural or extensional or shear-like.
Second, there are no universally accepted elastodynamic equations for curved structures, and in the case of the rod and the shell, several choices exist~\cite{waltking1934,morley1961,pierce1993,doyle2021,kernes2021}.
The simplest ones, however, are almost always a set of linear partial differential equations that couple the normal and tangential displacements, the independent variables being time and the coordinates that describe the undeformed configuration of structure ($x$ and $y$ in Fig.~\ref{fig:waves}).
In models that assume that the wavelength is larger than the thickness of the structure (assumed to be of order unity), this generally requires that the wavenumber $k$ and curvature $m$ satisfy the condition $0 \leq \abs{m} < \abs{k} \ll 1$~\cite{landau1986,pierce1993,pierce1993a,norris1995}.
Although not all elastodynamic models assume this condition~\cite{rose2010}, in this paper, we only consider ones that satisfy it.

Because the curvature couples the different displacement components, irrespective of the equations we use, to fully characterize wave propagation on curved structures, we have to consider multicomponent (i.e., vector) waves.
Computing the exact spectrum of a multicomponent differential operator is often difficult, unless one resorts to numerical or asymptotic techniques.
Indeed, for this reason, in their theoretical analyses of the musical saw, both \citet{scott1992}, and \citet{shankar2022} chose to simplify matters by analyzing flexural vibrations alone.

Most nontrivial wave problems can only be solved asymptotically, and the exact choice of the method depends on the problem at hand, e.g., in Refs.~\cite{gridin2005,gridin2005a} the authors resorted to an asymptotic method that assumes that wavelength is the same order as the inhomogeneity scale~\cite{adamou2005}.
In comparison, the semiclassical WKB (Wentzel--Kramers--Brillouin) approximation, widely applied in quantum-mechanical contexts, assumes that the wavelength is much smaller than the inhomogeneity scale.
It has also been extensively used to study elastic waves, which is not surprising given the many similarities between quantal and elastodynamic problems, including those describing the dynamics of thin structures such as plates and rods~\cite{bogomolny1998,lopez-gonzalez2021,mohammed2021,engstrom2023}.
When bound states are present, semiclassical asymptotics allows one to extract the corresponding frequencies through quantization~\cite{berry1972}.
Such an endeavor, however, becomes nontrivial in the case of multicomponent wave equations, such as the ones we study in this paper.
Here, subtleties can arise owing to the presence of an extra phase in the quantization rule~\cite{yabana1986,kaufman1987,littlejohn1991,littlejohn1991a}.
Recently, this phase has been shown~\cite{venaille2023} to be responsible for a spectral flow in the rotating shallow-water equations that describe oceanic waves on the Earth's surface, leading to the topological protection of equatorial waves~\cite{delplace2017}.

\subsection{Overview}

In this paper, we use the semiclassical approximation (summarized in Sec.~\ref{sec:wkb}) to study the bound-state spectrum of elastic waves in a curved rod and a singly curved shell with a varying curvature profile.
To avoid losing the main results of the paper in a thicket of details, we summarize them here:
\begin{enumerate}
  \item For both the rod and the shell, independent of the boundary conditions, waves exhibit robust localization around points where the absolute curvature has a minimum.
    Wave localization induced by the presence of an inflection point in an $\mathsf{S}$-shaped musical saw~\cite{scott1992,shankar2022} is a special case of this more general observation.
  \item In a curved rod, only extensional waves form bound states and flexural waves always form ``unbound'' states that are spread across the rod.
    We show this by using both semiclassical asymptotics (Sec.~\ref{sec:rods}) and by explicitly finding extensional solutions to the rod equations (Sec.~\ref{sec:extensional_limit}).
  \item In a shell, waves of all three types can form states that are bound along the curved direction
    [$x$ in Fig.~\ref{fig:waves}(b)].
    In the frequency spectrum, flexural bound states appear first and have the lowest frequencies.
    They are then followed by shear and extensional bound states, in that order (Sec.~\ref{sec:shell}).
  \item For both structures, flexural waves start propagating well below the frequency of the first bound state associated with an extensional wave.
    Hence, in very long rods and shells, these bound states coexist with a near-continuum of flexural waves, forming quasi-bound states in a continuum~\cite{hsu2016}.
  \item
    Finally, both structures are described by equations for which the extra phase in the modified quantization rule vanish---something that we expect to be generically true for equations of thin-walled structures.
    This simplifies our analysis considerably and results in remarkable agreement between our numerical experiments and quantization results.
\end{enumerate}

Our findings show that waves can be robustly trapped in thin elastic structures by a simple alteration of their geometry.
This could help, for instance, in crafting better thin-plate acoustic cloaks~\cite{farhat2009}, and aid the control of noise and vibration in thin structures~\cite{mace1987,hansen2012}.
Curvature-induced localization of waves could also be used to improve the acoustic black-hole effect in thin-walled structures~\cite{lee2017,pelat2020}, which at the moment relies primarily on wave localization caused by a power-law tapered thickness profile~\cite{krylov2020}.
Finally, a singly curved shell serves as a simple, yet effective single-mode waveguide that can steer flexural waves of specific frequencies in the uncurved direction.

\paragraph*{Notation.} Matrices are set in sans-serif type, e.g., $\mathsf{D}$. Operators are distinguished with a hat, e.g., $\hat{k}$.  Hats are dropped on operator ``symbols'', e.g., $k$.  Unless explicitly indicated otherwise, repeated indices are to be summed over.

\section{Semiclassical theory of waves}
\label{sec:wkb}

In this section we present a quick rundown of the semiclassical approximation as applied to multicomponent waves.
For more detailed descriptions, we refer to the book by \citet{tracy2014} and references therein.
We begin with a wave equation of the form
\begin{equation}
  \partial_{t}^{2}\Psi(x,t) + \widehat{\mathsf{H}}\Psi(x,t) = 0,
  \label{eq:full_wave_eq}
\end{equation}
where $\Psi(x,t)$ is an $N$-component wave field described by a one-dimensional coordinate $x$ and time $t$.
In elastodynamics, $\Psi$ is usually composed of displacements, e.g., for the rod we have $\Psi = (\zeta, u)$, and for the shell we have $\Psi = (\zeta, u, v)$ [see Figs.~\ref{fig:waves}(a) and~\ref{fig:waves}(b)].
Also, $\widehat{\mathsf{H}}$ is taken to be a Hermitian operator in the form of an $N\times N$ matrix, composed solely of spatial derivatives (i.e., powers of $\partial_{x}$) with time-independent coefficients.
Assuming that the waves are time harmonic with frequency $\omega$, i.e., $\Psi(x, t) = \psi(x)e^{\pm i\omega t}$, where $\psi(x)$ is the time-independent part of the wave field, Eq.~\eqref{eq:full_wave_eq} can be recast as
\begin{equation}
  \widehat{\mathsf{D}}\psi = 0,\quad \text{with}\enspace \widehat{\mathsf{D}} = \widehat{\mathsf{H}} - \omega^{2}\mathsf{I}_{N},
  \label{eq:ev_problem}
\end{equation}
where $\mathsf{I}_{N}$ is the $N\times N$ identity matrix.
If the coefficients of the spatial derivatives that appear in $\widehat{\mathsf{D}}$ are constants, then the eigenmodes $\psi$ are plain waves.
In what follows we assume that these coefficients are slowly varying, with the variation controlled by a single positive parameter $\epsilon \ll 1$.
It is useful to treat $\epsilon$ as an ordering parameter so that we can look for solutions at various orders of $\epsilon$.
To this end, we rescale $x \to \epsilon^{-1}x$ so that a derivative $\partial_{x}$ becomes $\epsilon \partial x$.
With analogy to quantum mechanics, this allows us to recast the derivatives in $\widehat{\mathsf{D}}$ in terms of the wavenumber/momentum operator $\hat{k} = -i\epsilon \partial_{x}$, with $\epsilon$ playing the role of Planck's constant.
Since we shall be considering $\widehat{\mathsf{D}}$ in the coordinate representation, the position operator $\hat{x} = x$.

We look for \emph{eikonal} solutions to Eq.~\eqref{eq:ev_problem} of the form $\psi(x) = A(x)e^{iS(x)/\epsilon}$, where the amplitude $A(x)$ is an $N$-component spinor with complex components, and $S(x)$ is a rapidly varying phase, playing the role of an action.
To solve Eq.~\eqref{eq:ev_problem} at various orders of $\epsilon$, it is convenient to make use of Weyl calculus, which allows one to map differential operators that are functions of $\hat{x}$ and $\hat{k}$ to ordinary functions, called Weyl symbols, defined on an $x$-$k$ phase space, and vice versa~\cite{chaichian2001,cohen2012}.
Functions of $x$ and $\hat{k}$ alone, e.g., $f(x)$ and $g(\hat{k})$, follow the straightforward mapping
\begin{equation}
  f(x) \to f(x),\enspace
  g(\hat{k}) \to g(k).
\end{equation}
The Weyl symbol of the product of two operators in terms of their individual symbols, however, is given by the Moyal formula~\cite{tracy2014}
\begin{equation}
  \begin{aligned}
    f(x)g(\hat{k}) &\to f(x)g(k) + \frac{i\epsilon}{2}\left\{f(x), g(k)\right\} + \mathcal{O}(\epsilon^{2})\\
                   &= f(x)g(k) + \frac{i\epsilon}{2}f'(x)g'(k) + \mathcal{O}(\epsilon^{2}).
  \end{aligned}
  \label{eq:weylrules}
\end{equation}
Above, $\{\cdot,\cdot\}$ is the $x$-$k$ Poisson bracket.

Converting each entry of the matrix operator $\widehat{\mathsf{D}}$ into a Weyl symbol, we get the $N\times N$ dispersion matrix $\mathsf{D}$, which we express in various orders of $\epsilon$ as $\mathsf{D} = \mathsf{D}^{(0)} + \epsilon\mathsf{D}^{(1)} + \mathcal{O}(\epsilon^{2})$.
Employing the eikonal ansatz, at $\mathcal{O}(\epsilon^{0})$, we find the matrix equation $\mathsf{D}^{(0)}A = 0$.
To satisfy this equation, at least one of the $N$ eigenvalues of $\mathsf{D}^{(0)}$, say $\lambda(x, k;\, \omega)$, must vanish so that $\det \mathsf{D}^{(0)}(x, k; \omega) = 0$.
A vanishing eigenvalue $\lambda$ and the associated normalized eigenvector $\tau$ describe different wave types or ``polarizations'' represented by Eq.~\eqref{eq:ev_problem}.
By a polarization we mean a linear subspace of the total wave field that is usually of a distinct physical nature, e.g., flexural waves on a curved rod.

Vanishing eigenvalues of $\mathsf{D}^{(0)}$ serve as the \emph{ray Hamiltonian} of a wave of a specific polarization.
 This leads us to the phase-space representation of waves as rays that satisfy the Hamilton's~equations
\begin{equation}
  \begin{alignedat}{2}
    \dot{x} &= \partial_{k} \lambda(x, k;\, \omega) &= \left\{x, \lambda\right\},\\
    \dot{k} &= -\partial_{x} \lambda(x, k;\, \omega) &= \left\{k, \lambda\right\},
  \end{alignedat}
  \label{eq:phase_space}
\end{equation}
where the overdot denotes derivatives with respect to a parameter that parameterizes the rays in the phase space.
As the $x$-$k$ phase space is two-dimensional, the solutions to Eq.~\eqref{eq:phase_space} are identical to the level curves defined by $\lambda(x, k; \omega) = 0$.
It should be noted that the phase-space representation breaks down when more than one eigenvalue of $\mathsf{D}^{(0)}$ simultaneously vanish, which leads to mode conversion between the different polarizations~\cite{tracy2014}.
However, as we shall see, mode conversion effects can safely be ignored for problems we study in this paper.

\subsection*{Bound waves in phase space}

\begin{figure}
  \begin{center}
    \includegraphics{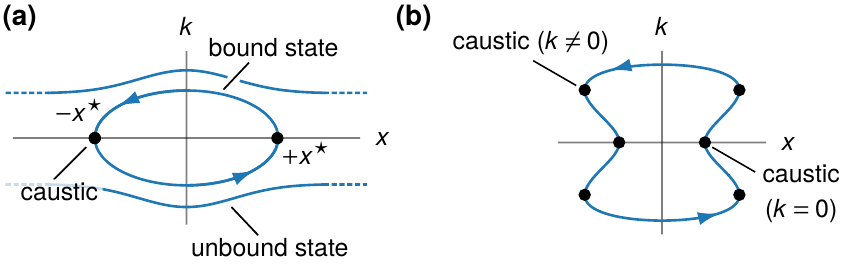}
  \end{center}
  \caption{%
    (a) In phase space, bound states are represented by rays in the form of closed orbits, which is analogous to that of a bound particle oscillating between two classical turning points ($\pm x^{\star}$ in the cartoon).
    Other trajectories represent unbound states.
    (b) A bound state represented by a ``peanut''-shaped orbit has six caustics.%
  }
  \label{fig:caustic}
\end{figure}

We expect the rays of bound waves to be bounded in phase space as well, with these rays being topologically equivalent to a circle~\cite{keller1958,mcdonald1988}.
As shown in Fig.~\ref{fig:caustic}(a), such rays oscillate between two classical turning points where $k = 0$ and $\dot{x} = 0$.
Turning points are examples of caustics, i.e., points on the ray where $\dot{x} = 0$, and in a bound ray, apart from the classical turning points, there could be other caustics as well [see Fig.~\ref{fig:caustic}(b)].
Even though the semiclassical approximation breaks down near the caustics, we can recover the phase $S(x)$ by integrating $k(x)$ along a ray.
Furthermore, for bound rays, single valuedness of $\psi(x)$ results in the modified Bohr--Sommerfeld quantization condition
\begin{equation}
  \epsilon^{-1}\oint \dd{x}\,k(x;\, \omega) = 2\left(n + \frac{\alpha}{4}\right)\pi - \gamma,
  \label{eq:quantization}
\end{equation}
from which bound-state frequencies can be obtained.
Above, the quantum number $n \in \mathbb{N}_{0}$ and $\alpha$ is the Keller--Maslov index~\cite{keller1958,maslov1981}.
Closed orbits in a two-dimensional phase space that can be smoothly deformed to a small circle always have $\alpha = 2$~\cite{percival1977}.
The additional phase $\gamma$ only appears when the wave field has more than one component, and is a consequence of the fact that the polarization vector $\tau$ is uniquely determined only up to an overall phase.
Its rate of change $\dot{\gamma}$ as we move along a ray can be written as $\dot{\gamma} = \dot{\gamma}_{\text{G}} + \dot{\gamma}_{\text{NG}}$, with~\cite{yabana1986,kaufman1987,venaille2023}
\begin{equation}
  \begin{aligned}
    \dot{\gamma}_{\text{G}} &= i\tau_{\mu}^{*}\left\{\tau_{\mu}, \lambda\right\} = i{\tau_{\mu}}^{\dagger}\dot{\tau}_{\mu},\\
    \dot{\gamma}_{\text{NG}} &= (i/2)\mathsf{D}^{(0)}_{\mu\nu}\left\{\tau^{*}_{\mu}, \tau_{\nu}\right\} - \tau_{\mu}^{*}\mathsf{D}^{(1)}_{\mu\nu}\tau_{\nu},
  \end{aligned}
\label{eq:extra_phases}
\end{equation}
where the asterisk represents complex conjugation and the subscripts $\mu, \nu$ represent the entries and components of $\mathsf{D}^{(0)}$ and $\tau$, respectively.
It can be shown that the first phase $\gamma_{\text{G}}$ has the general form of a geometric phase~\cite{pancharatnam1956,berry1984} upon treating the $x$-$k$ phase space as a parameter space~\cite{yabana1986}.
The second (nongeometric) phase $\gamma_{\text{NG}}$ has no such interpretation.

Instead of explicitly accounting for the extra phase $\gamma$ in the quantization rule, we could have diagonalized the wave equation at various orders of $\epsilon$~\cite{littlejohn1991,littlejohn1991a,weigert1993,venaille2023}.
During such a procedure, terms proportional to $\dot{\gamma}_{\text{G}}$ and $\dot{\gamma}_{\text{NG}}$ naturally appear in the ray Hamiltonian $\lambda$ as a first-order correction.
Despite the elegance of the method, we do not use it in our analysis.
This is because, as we discuss in Appendix~\ref{app:additional_phase}, for both the problems we consider, the extra phases vanish.

\section{Wave localization in a curved rod}
\label{sec:rods}

As we remarked earlier, several rod theories~\cite{chidamparam1993,walsh2000}, with varying levels of sophistication, have been written down to describe wave propagation on rods---straight or curved.
For our purposes, it is sufficient to work with a simple model~\cite{waltking1934,morley1961,graff1991,doyle2021} of a curved rod that ignores higher-order effects like torsion, cross-sectional rotation, etc.
We use this model when even more elementary rod theories exist (see, e.g., Appendix~\ref{app:rod_simple}) as it is more widely used, in addition to having received experimental attention~\cite{britton1968}.

\subsection{Equations of motion and semiclassical approximation}
\label{sec:rod_equations}

Let the undeformed state of the rod be in the form of a plane curve $\bm{\sigma}: \mathcal{X} \to \mathbb{R}^{2}$, which we take to be parameterized by its arclength $x \in \mathcal{X} \subset \mathbb{R}$.
As a wave propagates along the rod, it undergoes a deformation $\bm{\sigma} \to \bm{\sigma} + \delta\bm{\sigma}$, where the displacement field $\delta\bm{\sigma}(x,t) = u(x,t)\bm{t}(x) + \zeta(x,t)\bm{n}(x)$.
Here $\bm{t} = \dd\bm{\sigma}/\dd{x}$ is the unit tangent of the undeformed rod and $\bm{n}$ is its unit normal, obtained by rotating $\bm{t}$ counter-clockwise by $\pi/2$ [see Fig.~\ref{fig:waves}(a)].
We shall use the normal displacement $\zeta$ and the tangential displacement $u$ as the components of a two-component wave field $\Psi = (\zeta, u)$.
Assuming the absence of external forces, the equations of motion can be derived from the following action~\cite{doyle2021}:
\begin{equation}
  \begin{aligned}
    \mathscr{U}[\zeta, u] = \tfrac{1}{2}\int \dd{t}\,\dd{x}\,\Bigl[&\rho\left(\dot{u}^{2} + \dot{v}^{2}\right)- K\left(u' - m\zeta\right)^{2}\\
                                                         &- B\left(mu' + \zeta''\right)^{2}\Bigr],
  \end{aligned}
  \label{eq:rod_functional}
\end{equation}
where the overdot represents derivatives with respect to time $t$ and the primes denotes derivatives with respect to the arclength $x$.
We have also assumed that the rod is uniform with linear mass density $\rho$, with extensional stiffness $K$ and bending stiffness $B$.
Also, the signed curvature of the rod is $m(x) = \bm{n}\cdot\dd\bm{t}/\dd{x}$.
With these identifications, we can delineate the bending and stretching contributions to the energy.

In Eq.~\eqref{eq:rod_functional}, the extensional stiffness $K = YA$ and the bending stiffness $B = YI$, with $Y$ being the Young's modulus, $A$ being the cross-sectional area, and $I$ being the second moment of area.
If the rod's cross-sectional ``thickness'' is $h$, then $A \sim h^2$ and $I \sim h^{4}$.
This gives us a natural length unit $\ell = \sqrt{B/K} \sim h$ and a time unit $\sqrt{B\rho}/K$ that can be conveniently used for nondimensionalization. [Note that the curvature transforms as $m(x) \to \ell^{-1}m(x)$.]
Unlike Refs.~\cite{waltking1934,morley1961,graff1991,doyle2021}, where the curvature $m$ is assumed to be a constant, we will assume it varies with the arclength $x$.
Finally, upon varying the functional in Eq.~\eqref{eq:rod_functional}, we get the dynamic rod equations in the following nondimensional form [cf. Eq.~\eqref{eq:full_wave_eq}]:
\begin{widetext}
\begin{equation}
\partial_{t}^{2}
\begin{pmatrix}
  \zeta\\
  u
\end{pmatrix} +
\widehat{\mathsf{H}}
\begin{pmatrix}
  \zeta\\
  u
\end{pmatrix} = 0,
\enspace
\text{where}
\enspace
\widehat{\mathsf{H}}
=
\begin{pmatrix}
  \partial_{x}^{4} + m^{2} & -m\partial_{x}(1 - \partial_{x}^{2}) + 2m'\partial_{x}^{2} + m''\partial_{x}\\
  m\partial_{x}(1 - \partial_{x}^{2}) + m'(1 - \partial_{x}^{2}) & -(1 + m^{2})\partial_{x}^{2} - 2mm'\partial_{x}
\end{pmatrix}.
\label{eq:rod}
\end{equation}
Equation~\eqref{eq:rod} is a set of coupled equations involving the curvature $m(x)$ as the only parameter.
When $m = 0$, it decouples into two equations: the first of which is the Euler--Bernoulli beam equation that describes flexural vibrations that bend the rod and involves only $\zeta$; the second equation, which involves only $u$, characterizes extensional waves that propagate longitudinally by stretching the rod.
When curvature is nonzero, which is the case we want to analyze, the components $\zeta$ and $u$ remain coupled.

Because we have assumed that curvature $m(x)$ is nonuniform in Eq.~\eqref{eq:rod}, we can no longer seek a plane-wave solution.
Instead, we shall employ the semiclassical approximation.
To this end, we assume that the curvature is a slowly varying function of the form $m(\epsilon x)$.
Here $0 < \epsilon \ll 1$ is a small dimensionless parameter that controls the slowness of the variation.
Because the length unit $\ell$ we chose for nondimensionalizing the arclength $x$ is proportional to the thickness, physically speaking, here we are assuming that the length scale over which the curvature varies significantly is much larger that the thickness of the rod.
Next, we do a final change of variables $x \to \epsilon^{-1}x$ in Eq.~\eqref{eq:rod} so that $m(\epsilon x) \to m(x)$ and all spatial derivatives get multiplied by $\epsilon$.
For the eigenvalue problem with $\widehat{\mathsf{D}} = \widehat{\mathsf{H}} - \omega^{2}\mathsf{I}_{2}$, this gives us
\begin{equation}
  \begin{aligned}
    \widehat{\mathsf{D}} &=
    \begin{pmatrix}
      \epsilon^{4}\partial_{x}^{4} + m^{2} & -\epsilon m \partial_{x}(1 - \epsilon^{2}\partial_{x}^{2}) + 2\epsilon^{3}m'\partial_{x}^{2} + \epsilon^{3}m''\partial_{x}\\
      \epsilon m\partial_{x}(1 - \epsilon^{2}\partial_{x}^{2}) + \epsilon m'(1 - \epsilon^{2}\partial_{x}^{2}) & -(1 + m^{2})\epsilon^{2}\partial_{x}^{2} - 2\epsilon^{2}mm'\partial_{x}
    \end{pmatrix}\\
                         &=
                         \begin{pmatrix}
                           \hat{k}^{4} + m^{2} -\omega^{2} & -i m\hat{k}(1+\hat{k}^{2}) - 2\epsilon m'\hat{k}^{2} + i\epsilon^{2}m''\hat{k}\\
                           im\hat{k}(1+\hat{k}^{2}) + \epsilon m' (1+\hat{k}^{2}) & (1+m^{2})\hat{k}^{2} - 2i\epsilon mm'\hat{k}
                         \end{pmatrix}.
  \end{aligned}
  \label{eq:filwaveop}
\end{equation}
In the last step above, we have set $\epsilon\partial_{x} \to i\hat{k}$, with $\hat{k}$ being the ``momentum'' operator.
Using the rules in Eq.~\eqref{eq:weylrules} we can easily write down the Weyl symbol $\mathsf{D}$ for the operator in Eq.~\eqref{eq:filwaveop} as
$\mathsf{D} \approx \mathsf{D}^{(0)} + \epsilon\mathsf{D}^{(1)}$,
where
\begin{equation}
    \mathsf{D}^{(0)} =
    \begin{pmatrix}
      {k}^{4} + m^{2} - \omega^{2} & -i mk(1 + k^{2})\\
      im{k}(1 + k^{2}) & (1+m^{2}){k}^{2} - \omega^{2}
    \end{pmatrix}
    \quad\text{and}\quad
    \mathsf{D}^{(1)} =
    \tfrac{1}{2}
    \begin{pmatrix}
      0 & m'(1-k^{2})\\
      m'(1-k^{2}) & 0
    \end{pmatrix}.
  \label{eq:rod_D}
\end{equation}
Here we have ignored the $\mathcal{O}(\epsilon^{2})$ correction $\mathsf{D}^{(2)}$, which would be superfluous to include in the first-order eikonal approximation we use in this paper (see Sec.~\ref{sec:wkb}).
The two eigenvalues of the lowest-order dispersion matrix $\mathsf{D}^{(0)}$, representing waves of two different polarizations,~are
\begin{equation}
    \lambda_{\pm}(x, k; \omega) = \tfrac{1}{2}
  \left\{(1+k^{2})\left[k^{2}+m^{2}(x)\right] \pm \sqrt{\left[k^{2}-m^{2}(x)\right]^{2}(1-k^{2})^{2} + 4m^{2}k^{2}(1+k^{2})^{2}}\right\} - \omega^{2}.
  \label{eq:rod_ham}
\end{equation}
\end{widetext}
For a given $\omega$, we have $\lambda_{+}(x, k;\, \omega) \geq \lambda_{-}(x, k;\, \omega)$ for all values of $x$ and $k$.
Mode conversion between the two polarizations ensues near points where $\lambda_{+}(x, k;\, \omega) = \lambda_{-}(x, k;\, \omega)$, and the semiclassical approximation breaks down.
We see that $\lambda_{+} = \lambda_{-}$ only when the discriminant in Eq.~\eqref{eq:rod_ham} vanishes, which happens only when $m(x)$ is zero, and $k = \pm 1$ or $k = 0$.
As we discussed in Sec.~\ref{sec:intro_background}, the rod equations are only applicable for short waves whose wavelength is much longer than the thickness, which translates to the requirement $0 \ll \abs{k} \ll 1$.
Hence, mode-conversion points with $k = \pm 1$ or $k = 0$, lie well beyond the range of applicability of these equations.
For this reason, we ignore mode-conversion issues and assume that $\lambda_{+} \neq \lambda_{-}$ throughout our analysis.
Before moving on, it is useful to first analyze the propagation of waves on a rod of constant curvature.

\subsection{Rods of constant curvature}

\begin{figure*}
  \begin{center}
    \includegraphics{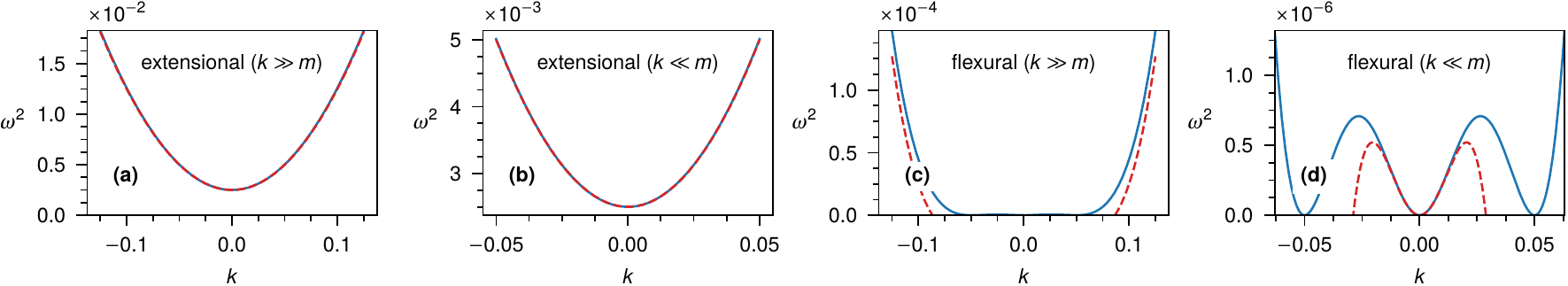}
  \end{center}
  \caption{Dispersion curves of plane waves propagating on a rod of constant curvature $m = 0.05$.
    The solid blue curves represent the exact dispersion relations obtained after setting $\lambda_{\pm} = 0$ in Eq.~\eqref{eq:rod_ham}.
    The red dashed curves represent the asymptotic dispersion relations in the $k \gg m$ and $k \ll m$ limits and given in Eqs.~\eqref{eq:rod_kgm} and \eqref{eq:rod_klm}, respectively.}
    \label{fig:rod_dispersion}
\end{figure*}

First we analyze the case where the curvature is zero, i.e., when the rod is straight.
In the limit of vanishing curvature $m$, we should recover the dispersion relations $\omega(k)$ for plane waves propagating on a straight rod from $\lambda_{\pm}$.
On setting $m = 0$ and noting that $\abs{k} \ll 1$, we see that $\lambda_{+} = 0$ gives us the linear dispersion relation $\omega = k$, representing extensional waves propagating on a straight rod.
Meanwhile, when $m = 0$, we find that $\lambda_{-} = 0$ gives us the quadratic dispersion relation $\omega = k^{2}$ of flexural waves on a straight rod.

For nonzero, but constant curvature, the rod forms part of a ring.
If the curvature is sufficiently weak, we expect the eigenvalue $\lambda_{+}$ to continue to represent predominantly extensional waves and $\lambda_{-}$ to represent predominantly flexural waves.
To verify this, we expand $\lambda_{\pm}$ in powers of $m^{2}$ and drop powers of $k$ in comparison to unity to find
\begin{equation}
  \begin{aligned}
  \begin{aligned}
    \lambda_{+} &= k^{2} + m^{2} - \omega^{2} + \mathcal{O}(m^{4}),\\
    \lambda_{-} &= k^{4} - 3k^{2}m^{2} - \omega^{2} + \mathcal{O}(m^{4})
  \end{aligned}
  & \qquad{(k \gg m)}.
  \label{eq:rod_kgm}
  \end{aligned}
\end{equation}
Clearly, the above expansions can only be valid when the $\mathcal{O}(m^{2})$ correction terms are less than the lowest-order terms, which is true only when $k \gg m$.
The asymptotic dispersion relations obtained from Eq.~\eqref{eq:rod_kgm} are illustrated in Figs.~\ref{fig:rod_dispersion}(a) and \ref{fig:rod_dispersion}(c), respectively.

We next look at the case where both $k$ and $m$ are small, but with $k \ll m$.
We consider this limit to later analyze the behavior of waves close to a classical turning point where $k = 0$.
Such waves decay beyond the turning point, and they never get a chance to complete a full-wavelength oscillation with $k \ll m$, so the short wavelength assumption, namely that the wavelength is shorter than the radius of curvature~\cite{pierce1993}, is not violated.
To consider the limit $k \ll m$, we expand $\lambda_{\pm}$ in powers of $km^{-1}$ and find
\begin{equation}
  \begin{aligned}
  \begin{aligned}
    \lambda_{+} &= k^{2} + m^{2} - \omega^{2} + \mathcal{O}(k^{4}),\\
    \lambda_{-} &= k^{2}m^{2} - 3k^{4}\left[1 + \mathcal{O}\left(\frac{k^{2}}{m^{2}}\right)\right] - \omega^{2}
  \end{aligned}
  &\qquad{(k \ll m)}.
  \end{aligned}
  \label{eq:rod_klm}
\end{equation}
Dispersion relations obtained from setting $\lambda_{\pm} = 0$ above show deviation compared to Eq.~\eqref{eq:rod_kgm}, especially in the case of flexural waves [see Figs.~\ref{fig:rod_dispersion}(b) and \ref{fig:rod_dispersion}(d)].
For the same reason, in this limit, we expect waves of both polarizations to have both longitudinal and transverse characteristics.
Despite this, for the sake of simplicity and identification, we shall continue to call waves represented by $\lambda_{+}$ as extensional waves and those represented by $\lambda_{-}$ as flexural waves.

The bending of the rod is associated with the normal component $\zeta$ and stretching with the tangential component $u$.
To better understand how the two components contribute to the wave field in the presence of curvature, we define the amplitude ratio
\begin{equation}
  \mathscr{R} = \frac{\abs{\zeta}}{\abs{\zeta} + \abs{u}}.
  \label{eq:rod_ratio}
\end{equation}
With the above definition, for purely transverse and longitudinal waves $\mathscr{R} = 1$ and $\mathscr{R} = 0$, respectively.
In the eikonal ansatz, at the lowest order, the wave field $\psi = (\zeta, u)$ is proportional to the eigenvectors $\tau_{\pm}$ of $\mathsf{D}^{(0)}$, and so $\zeta \sim \tau_{\pm,1}$ and $u \sim \tau_{\pm,2}$.
Making use of Eqs.~\eqref{eq:rod_kgm} and \eqref{eq:rod_klm}, to the lowest order in $k$ and $m$, the eigenvectors $\tau_{\pm}$ are
\begin{equation}
  \tau_{+} \sim
  \begin{pmatrix}
    m\\
    ik
  \end{pmatrix}
  \quad\text{and}\quad
  \tau_{-} \sim
  \begin{pmatrix}
    ik\\
    m
  \end{pmatrix},
  \label{eq:rod_tau}
\end{equation}
so the asymptotic amplitude ratios for the two wave polarizations become $\mathscr{R}_{+} \sim \abs{m}/(\abs{m} + \abs{k})$ and $\mathscr{R}_{-} \sim \abs{k}/(\abs{m} + \abs{k})$.
For $m = 0$, we know that extensional waves become entirely longitudinal, and as expected the corresponding amplitude ratio $\mathscr{R}_{+} = 0$.
In same limit, flexural waves become entirely transverse as indicated by $\mathscr{R}_{-} = 1$.
However, for small values of $k$ with $k \ll m$ we see that $\mathscr{R}_{+} \to 1$ and $\mathscr{R}_{-} \to 0$.
In other words, in the limit $k \ll m$, waves of the two polarizations would switch their nature from being predominantly longitudinal to being predominantly transverse, and vice versa.

\subsection{Rods with varying curvature}

\begin{figure}
  \begin{center}
    \includegraphics{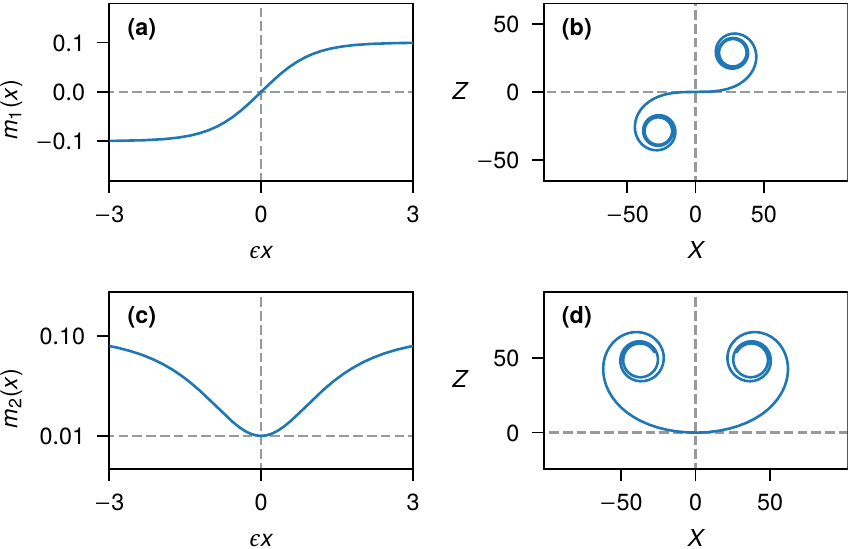}
  \end{center}
  \caption{%
    (a) Tanh-type curvature profile $m_{1}(x)$ with an inflection point at $x = 0$ and (b) the corresponding shape of the rod in Cartesian space with coordinates $X$ and $Z$.
    (c) Sech-type curvature profile $m_{2}(x)$ with no inflection point and (d) the corresponding shape of the rod.
    For both curvature types, as the arclength $x \to \pm\infty$, the rod becomes part of a circle.
    In all figures $a = 0.01$, $b = 0.1$, and $\epsilon = 0.01$.
  }
  \label{fig:rod_profile}
\end{figure}

\begin{figure*}
  \begin{center}
    \includegraphics{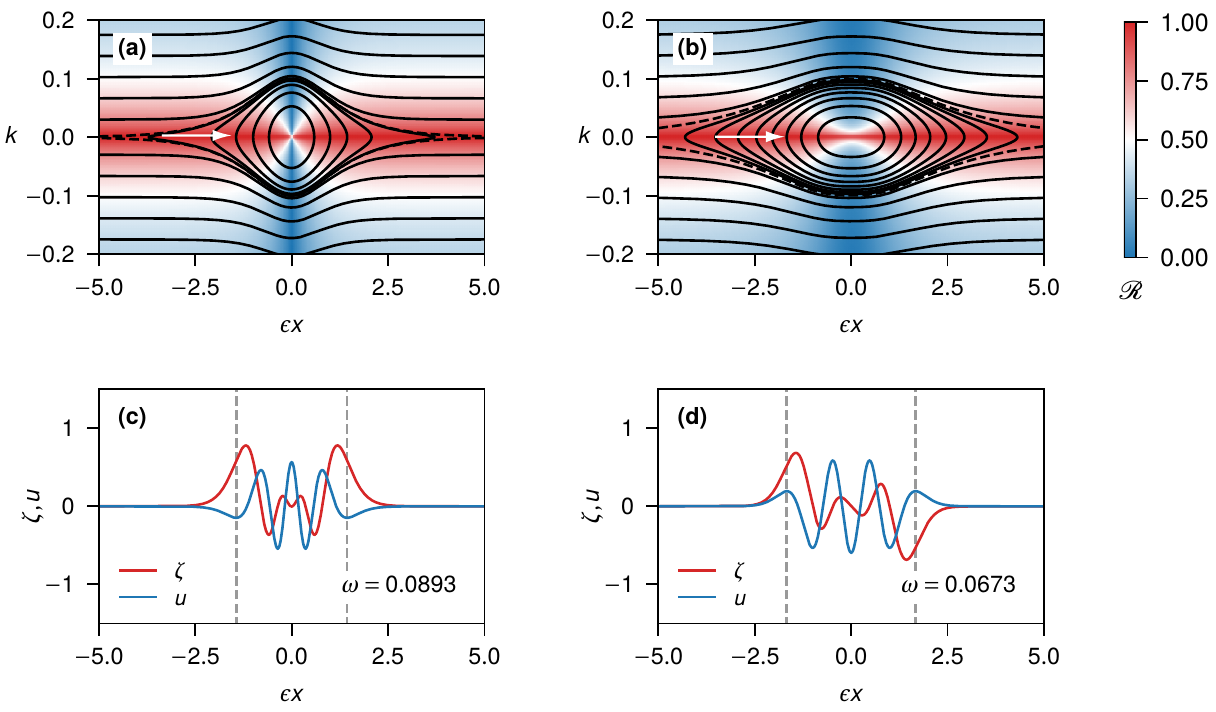}
  \end{center}
  \caption{%
    Ray trajectories for extensional waves on a rod with (a) tanh-type curvature profile $m_{1}(x)$ and (b) sech-type curvature profile $m_{2}(x)$, with the phase portraits color coded using the amplitude ratio $\mathscr{R}$ defined in Eq.~\eqref{eq:rod_ratio}.
    The white arrows in panels (a) and (b) indicate rays of the same frequency as the two bound states in (c) and (d).
    The grey vertical lines in panels (c) and (d) indicate the locations of the classical turning points.
  }
  \label{fig:rod_bound}
\end{figure*}

For illustrative purposes, we consider rods of two curvature profiles $m_{1}(x)$ and $m_{2}(x)$, defined by
\begin{equation}
  \begin{aligned}
  m_{1}(x) &= b\tanh(\epsilon x),\\
  m_{2}(x) &= b - \left(b - a\right)\sech(\epsilon x),
  \end{aligned}
  \label{eq:rod_curv}
\end{equation}
which we will informally call tanh- and sech-type curvature profiles, respectively.
Above, $b$ and $a$ are positive constants such that $a < b$ and the parameter $\epsilon$ controls the slowness of variation of the curvatures.
A rod with a tanh-type curvature profile $m_{1}(x)$ possesses an inflection point at $x=0$ where the curvature vanishes, and the curvature asymptotes to $\pm b$ as $x\to \pm\infty$ [see Figs.~\ref{fig:rod_profile}(a) and~\ref{fig:rod_profile}(b)].
On the other hand, a rod with a sech-type curvature profile $m_{2}(x)$ remains concave for all $x$, with the curvature acquiring its minimum value of $a$ at $x = 0$ and asymptoting to $b$ as $x\to\pm\infty$ [see Figs.~\ref{fig:rod_bound}(c) and~\ref{fig:rod_bound}(d)].
For the purpose of illustrating our results, we take $b = 0.1$ and $a = 0.01$, so the smallest ratio between the radius of curvature and thickness is $\mathcal{O}(10)$.
Although such a ratio is comparatively smaller than typical experimental values, it would give us a clearer picture of the effects of a varying curvature profile.
We also set $\epsilon = 0.01$, but as we rescale $x \to \epsilon^{-1}x$, the parameter $\epsilon$ does not always explicitly appear in our discussions.
Note also that the numerical values of the local wavenumber $k$ is unaffected by this rescaling.

For a varying curvature profile, using expressions for $\lambda_{\pm}$ from Eq.~\eqref{eq:rod_hamiltonian}, we find the rays for both polarizations by integrating the corresponding Hamilton's equations given by
\begin{equation}
  \dot{x} = \frac{\partial \lambda_{\pm}}{\partial k}
  \quad
  \text{and}\quad
  \dot{k} = -\frac{\partial \lambda_{\pm}}{\partial x}.
  \label{eq:rod_hamiltonian}
\end{equation}
One can also find the rays graphically by simply plotting the level curves of $\lambda_{\pm}$ in the $x$-$k$ phase space---a specific value of $\omega$ corresponding to a specific ray.

\subsection{Extensional waves}
\label{sec:extensional}

We first discuss how a varying curvature profile affects extensional waves in a rod.
As we discussed previously, extensional waves are to be associated with the ray Hamiltonian $\lambda_{+}$.
Without loss of generality, let $x = 0$ be a point where the function $m^{2}(x)$ has a local extremum, so that its derivative $[m^{2}(x)]'\big|_{x=0} =~0$.
Then, straightforward algebra reveals that the origin $(0, 0)$ becomes a fixed point of Eq.~\eqref{eq:rod_hamiltonian} with $\lambda_{+}$ as the Hamiltonian.
Since Eq.~\eqref{eq:rod_hamiltonian} is a Hamiltonian system, the origin becomes a nonlinear center or a saddle depending on whether the Hamiltonian $\lambda_{+}(x, k)$, viewed as a function of $x$ and $k$, has an extremum or a saddle there~\cite{strogatz1994,jordan2007}.
For the present problem, we find two situations%
\footnote{We can also infer this from the sign of the Hessian determinant of $\lambda_{+}$ at $(0, 0)$ given by $\det \nabla\nabla \lambda_{+}\big|_{x=0,k=0} = 2(m^{2})''\big|_{x=0}$, provided that $(m^{2})'' \neq 0$ so that the origin is a nondegenerate fixed point with $\det\nabla\nabla\lambda_{+} \neq 0$.}
depending on the behavior of $m^{2}(x)$ at $x = 0$.
\begin{enumerate}
  \item \emph{$m^{2}$ has a nonzero maximum at $x = 0$.\enspace}
    Then, for any point $x$ sufficiently close to $0$, we have $m^{2}(x) < m^{2}(0) = \lambda_{+}(0, 0)$.
    We also see that $\lambda_{+}(x, 0) - \lambda_{+}(0, 0) < 0$, whereas $\lambda_{+}(0, k) - \lambda_{+}(0, 0) > 0$.
    In other words, if we move along the $x$ axis from the origin $(0, 0)$, the value of $\lambda_{+}(x, k)$ decreases, whereas moving along the $k$ axis increases its value, showing that the origin becomes a saddle point.
  \item \emph{$m^{2}$ has a minimum at $x = 0$.\enspace}
    At any point $(x, k)$ sufficiently close to $(0, 0)$, we have $m^{2}(x) > m^{2}(0)$.
    Then, using basic inequality arguments, $\lambda_{+}(x, k) - \lambda_{+}(0, 0) > 0$, showing that $\lambda_{+}$ has a minimum at the origin, which means that it becomes a nonlinear center.
\end{enumerate}
As the extrema of $m^{2}(x)$ are identical to the extrema of the absolute curvature $\abs{m(x)}$, we can summarize as follows: points where the absolute curvature has a minimum become nonlinear centers of the ray equations associated with $\lambda_{+}$, and points where the absolute curvature has a maximum become saddles.
Therefore, we expect extensional waves to get trapped only around points where the absolute curvature has a minimum.

The black curves in Figs.~\ref{fig:rod_bound}(a) and~\ref{fig:rod_bound}(b) show the extensional rays for the example curvature profiles.
Each ray is a level curve defined by $\lambda_{+}(x, k; \omega) = 0$ for a specific $\omega$.
Since $\lambda_{+}$ is invariant with respect to reflection about the $x$ and $k$ axes, the ray trajectories also share the same reflection symmetry.
For both curvature types, there are three fixed points: the origin $(0, 0)$ and $(\pm\infty, 0)$.
The fixed points at $(\pm\infty, 0)$ are saddles since $m^{2}(x)$ achieves its maximum there.
Two heteroclinic orbits, shown as black dashed curves, connect the fixed points at $(\pm\infty, 0)$.
As per our analysis, the origin is always a center as $m^{2}(x)$ achieves a minimum there for both curvature types.
Close to the origin, the rays appear in form of closed orbits indicating bound states.

Two example bound states obtained by solving the rod equations numerically (see Appendix \ref{app:numerical} for details) are shown in Figs.~\ref{fig:rod_bound}(c) and~\ref{fig:rod_bound}(d).
As we see from these figures, the components $\zeta(x)$ and $u(x)$ have the same parity for odd $m(x)$, but always have different parity when $m(x)$ is even.
This is not surprising.
If $\psi(x) = \left[\zeta(x), u(x)\right]$ is a bound state, then it is easy to check that for an odd curvature profile $m(x)$ with $m(-x) = -m(x)$, the eigenmode $\psi(-x) = \left[\zeta(-x), u(-x)\right]$ is also a bound state satisfying the rod equations, Eq.~\eqref{eq:rod}, with the boundary conditions $\psi(\pm\infty) = 0$.
Assuming nondegeneracy in the eigenmodes, we must then have $\left[\zeta(x), u(x)\right] = \pm\left[\zeta(-x), u(-x)\right]$, which means that for an odd $m(x)$, the components $\zeta(x)$ and $u(x)$ must have the same parity, i.e., they are either both odd or both even.
For an even $m(x)$, however, we find that bound-state solutions must satisfy $\left[\zeta(x), u(x)\right] = \pm\left[\zeta(-x), -u(-x)\right]$, showing that the components $\zeta(x)$ and $u(x)$ always have different parity.

The rays corresponding to the bound states in Figs.~\ref{fig:rod_bound}(c) and~\ref{fig:rod_bound}(d) have been marked with white arrows in the phase portraits in Figs.~\ref{fig:rod_bound}(a) and~\ref{fig:rod_bound}(b).
  Clearly, the displacement fields of the bound states of both curvature profiles show a significant presence of both normal and tangential components.
This can also be inferred from the phase portraits, which have been color coded using the amplitude ratio $\mathscr{R}$ defined in Eq.~\eqref{eq:rod_ratio}, and computed from the components of the polarization vector $\tau_{+}$.
Starting on the $k$ axis, if we move along a closed orbit towards one of its turning points, we see that $\mathscr{R}$ changes from $0$ to $1$, indicating that the wave acquires a strong normal component, as we see in the numerical examples.

The existence of the bound states can also be intuitively understood from the dispersion relations for extensional waves.
From Eq.~\eqref{eq:rod_ham}, we see that at $k = 0$, the dispersion curve given by $\lambda_{+} = 0$ has a nonzero gap and a cut-on frequency of $\omega_{\text{cut-on}}^{2} = m^{2}$.
For waves with $\omega < \omega_{\text{cut-on}}$, the wavenumber $k$ is always complex, preventing the wave from being able to propagate and the wave decays.
Now, as an extensional wave enters a region of high curvature from a region of low curvature, the local cut-on frequency increases.
This means that, at some point, the frequency of the wave would fall below the local cut-on frequency, and the wave gets reflected creating a bound state.
As the cut-on frequency depends only on the magnitude of the curvature, and not its sign, bound states do not require the presence of an inflection point---an observation that was also made by \citet{scott1992} while analyzing the musical saw.

An interesting analogy with geometric optics and the bound states is revealed if we consider the phase velocity $c_{m} = \omega/k$ of extensional waves on a curved rod with curvature $m$.
Upon defining an effective refractive index $c_{0}/c_{m}$ and using the dispersion relations in Eq.~\eqref{eq:rod_kgm}, extensional waves are seen to have a refractive index of $\sqrt{\omega^{2} - m^{2}}/\omega \sim 1 - m^{2}/(2\omega^{2})$, assuming $m\ll \omega$.
Thus, it would appear that the bound extensional waves seen in our examples are trapped inside regions of high refractive index (low curvature).
This is the working principle behind many acoustic black holes and insulators where variations in physical parameters result in a refractive index gradient, which in turn allows one to trap waves in high index regions~\cite{climente2013}.

\vspace{-0.5\baselineskip}
\subsubsection*{Bound state quantization}
\vspace{-0.5\baselineskip}

\begin{figure}
  \begin{center}
    \includegraphics{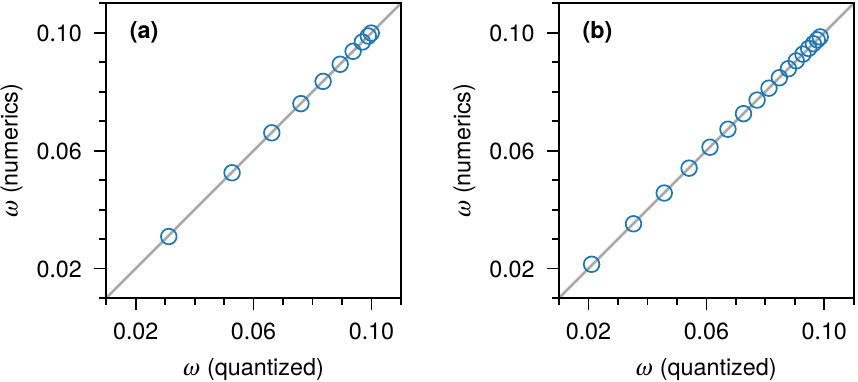}
  \end{center}
  \caption{
    Bound-state frequencies obtained from numerics compared to that obtained through quantization for a rod with (a) tanh-type curvature profile and (b) sech-type curvature profile.
    In both plots, the gray guideline in the background represents $\omega$ (numerics) = $\omega$ (quantized).
  }
  \label{fig:rod_wkb}
\end{figure}

To evaluate the action in the quantization condition, Eq.~\eqref{eq:quantization}, in principle, we should express $k$ in terms of $x$ from $\lambda_{+}(x, k; \omega) = 0$.
This proves to be difficult as $k$ would then have to be obtained as the root of a sixth-order polynomial.
Instead, as described in Appendix~\ref{app:numerical}, we compute the action by numerically integrating $k(x)$ between the classical turning points $\pm x^{\star}$.
Setting $k = 0$ in $\lambda_{+}(x, k; \omega) = 0$, we see that these points are implicitly given by the equation $m^{2}(x^{\star}) = \omega^{2}$.
To quantize the bound orbits, we also need the Keller--Maslov index, which is $\alpha = 2$ as the orbits are topologically equivalent to a circle.
Now, note that the off-diagonal entries of the dispersion matrix $\mathsf{D}^{(0)}$, Eq.~\eqref{eq:rod_D}, are purely imaginary.
In Appendix~\ref{app:additional_phase}, we show that the additional phase $\gamma$ in Eq.~\eqref{eq:quantization}, vanishes for any dispersion matrix of this form, enabling us to find the quantized frequencies without additional difficulty.
A comparison between the numerically obtained bound-state frequencies and those obtained through quantization (Fig.~\ref{fig:rod_wkb}) shows excellent agreement between the two.
Furthermore, these frequencies are independent of the boundary conditions chosen to solve the rod equations, showing the robustness of the bound states.

Instead of computing the action by quadrature, we could have extracted $k(x)$ from the approximate expression for $\lambda_{+}$ [see Eqs.~\eqref{eq:rod_kgm} and \eqref{eq:rod_klm}], which gives $k(x) \approx \pm\sqrt{\omega^{2} - m(x)^{2}}$.
This expression can be used, for instance, to estimate the maximum number of bound states $n_{\text{max}}$ for the two curvature profiles.
The largest frequency for which we see an extensional bound state is $\omega = b$.
As shown in Figs.~\ref{fig:rod_bound}(a) and~\ref{fig:rod_bound}(b), the rays corresponding to this frequency form two heteroclinic orbits connecting the classical turning points at $x = \pm \infty$.
Computing the area inside these orbits, we find
\begin{flalign}
  \hspace{-0.5em}
  n_{\text{max}} \approx \frac{1}{\pi\epsilon}\int_{-\infty}^{+\infty} \dd{x}\,k(x) =
  \begin{dcases}
    b/\epsilon & \text{(tanh type)},\\
    \bar{b}/\epsilon + \mathcal{O}(a/\epsilon) & \text{(sech type)}.
  \end{dcases}
  \label{eq:rod_maxb}
\end{flalign}
Here, $\bar{b} = \pi^{-1}b\int_{-\infty}^{\infty}\dd{x}\,\sech{x}\sqrt{2\cosh{x} - 1} \approx 2b$.
Therefore, we expect an infinitely long rod with a sech-type curvature profile to support twice as many bound states as a rod with a tanh-type curvature profile.
A similar expression for $n_\text{max}$ is also derived in the next section where we directly consider the extensional limit of the rod equations, and find the quantized frequencies as well as the bound modes.

\begin{figure*}
  \begin{center}
    \includegraphics{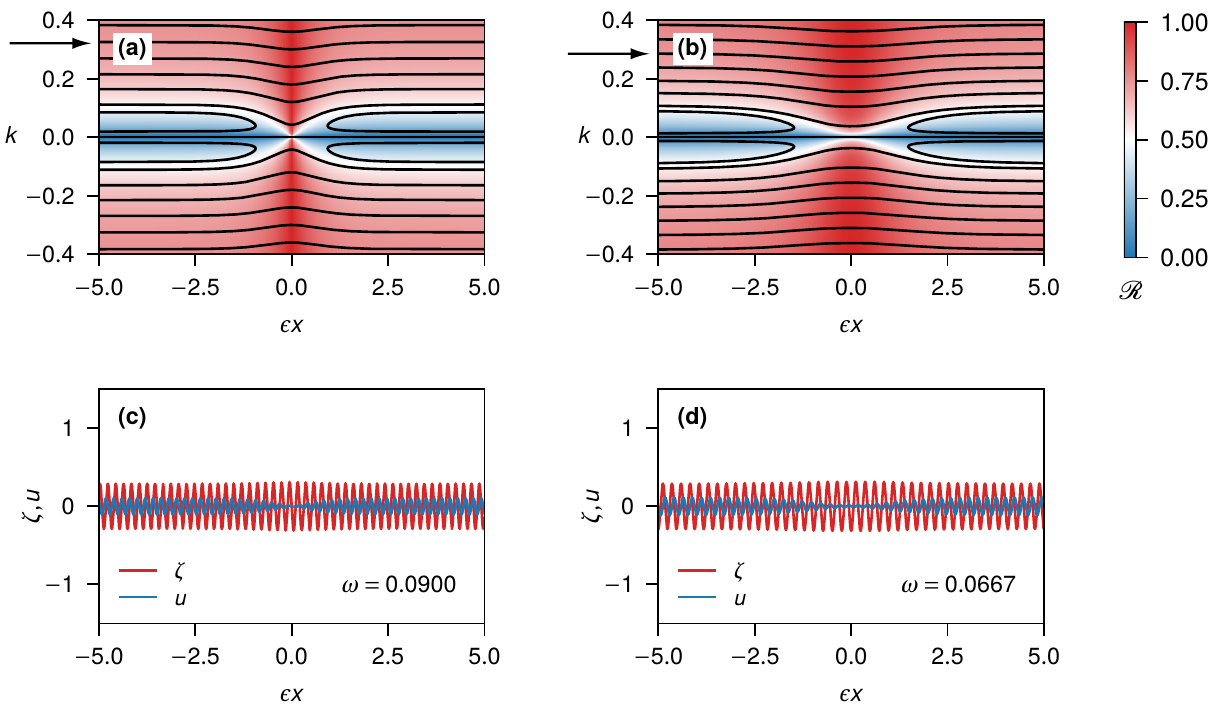}
  \end{center}
  \caption{%
    Ray trajectories for flexural waves on a rod with (a) tanh-type curvature profile $m_{1}(x)$ and (b) sech-type curvature profile $m_{2}(x)$, with the phase portraits color coded using the ratio $\mathscr{R}$ defined in Eq.~\eqref{eq:rod_ratio}.
    The black arrows in panels (a) and (b) indicate rays of the same frequency as the two eigenmodes in panels (c) and (d).
  }
  \label{fig:rod_extended}
\end{figure*}

For the values $b = 0.1$ and $\epsilon = 0.01$ we use in our examples, Eq.~\eqref{eq:rod_maxb} predicts a maximum of 10 bound states for a tanh-type rod and 20 bound states for a sech-type rod.
This prediction is consistent with our numerical experiments with a finite-sized rod, where we find a total of 10 and 17 bound states for the two curvature profiles.
Despite this, we only expect a small number of extensional bound states in actual experiments with curved rods, where we expect the finiteness of the rod to become more important.
For one thing, as the arc length $x \to \pm\infty$, self-intersection is inevitable for both sech- and tanh-type rods having a nonzero thickness (see Fig.~\ref{fig:rod_profile}).
(Self-intersection effects, however, can be ameliorated if we allow for small motions of the rod perpendicular to the plane containing it.)
Furthermore, as the turning points of the higher-frequency bound states are far apart, they would not always appear to be spatially localized, even though they decay exponentially as $x \to \pm\infty$.

It is natural to wonder if the bound modes we have observed are related to the more elementary vibrational modes seen in curved rings.
For instance, we see from the rod profiles in Fig.~\ref{fig:rod_profile} that for large $x$, the two ends of the rod spiral around a ring whose curvature is $b = 0.1$.
If we approximate the two ends of the rod as ideal rings, we see that the ring frequency at this curvature is just $b$ (extensional waves have unit speed in our conventions).
Waves at the ring frequency have a wavelength equal to the circumference $2\pi/b$ and beyond them lie other elementary modes of the ring such that the circumference is equal to integral multiples of the wavelength.
However, from the results in Fig.~\ref{fig:rod_wkb}, we see that all the bound states lie below these elementary modes and are independent of them, which is not so surprising given that two farthest classical turning points in phase space must be on or inside an orbit with frequency $\omega = b$.

\subsection{Flexural waves}

Flexural waves are associated with the ray Hamiltonian $\lambda_{-}$ and by inspecting the corresponding Hamilton's equations in Eq.~\eqref{eq:rod_hamiltonian}, we see that the derivatives $\dot{x}$ and $\dot{k}$ vanish everywhere on the $x$ axis.
Since there are no isolated fixed points anywhere on the $x$ axis for these rays, we do not expect flexural waves to form bound states.
Two example phase portraits for flexural waves are showcased in Figs.~\ref{fig:rod_extended}(a) and~\ref{fig:rod_extended}(b).
As we see from these figures, even though the rays do not form closed orbits, in regions where $k \ll m$, they have a nontrivial appearance.
This can be explained by the triple-well nature of the flexural dispersion curves at low $k$ [see Fig.~\ref{fig:rod_dispersion}(d)], which results in four separate rays for a single value of $\omega$.
That said, waves that lie entirely within the $k \ll m$ region have wavelengths larger than the local radius of curvature, and are beyond the range of applicability of the rod model we use.
For this reason, we forgo a more detailed analysis of rays with low wavenumbers.

Two unbound flexural eigenmodes obtained numerically are displayed in Figs.~\ref{fig:rod_extended}(c) and~\ref{fig:rod_extended}(d).
Rays corresponding to these modes have been marked by black arrows in Figs.~\ref{fig:rod_extended}(a) and~\ref{fig:rod_extended}(b).
Although the normal component $\zeta$ dominates in these modes, they acquire a significant tangential component with increasing curvature, which we also see from the color coding of the phase portraits.

\begin{figure*}
  \begin{center}
    \includegraphics{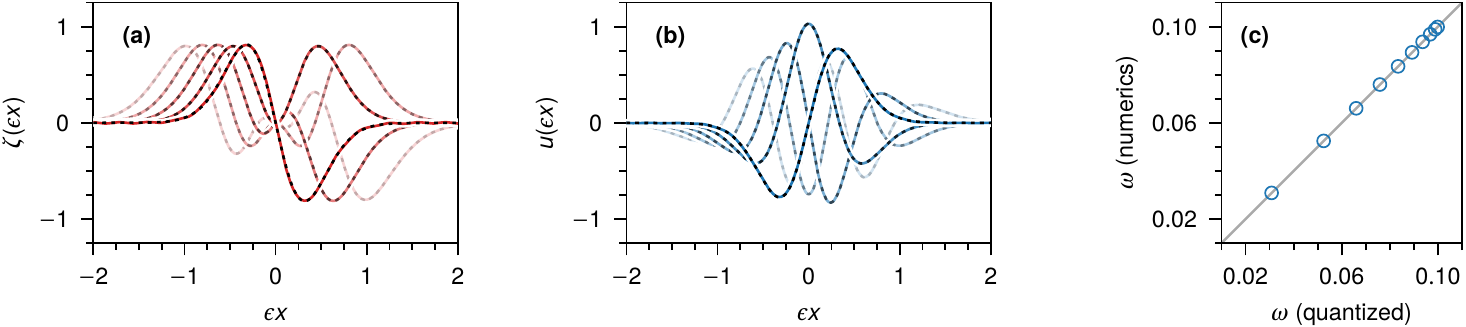}
  \end{center}
  \caption{%
    The first five extensional bound states of a rod with a tanh-type curvature profile showing (a) the normal component $\zeta(x)$ and (b) the tangential component $u(x)$, with lighter curves depicting higher-frequency states.
    The dashed black curves in panels (a) and (b) represent the solutions in Eq.~\eqref{app:eq:poschl}.
    (c) Comparison between the bound-state frequencies $\omega$ obtained through quantization, and those from numerics [cf.~Fig.~\ref{fig:rod_wkb}(a)].
  }
  \label{fig:poschl}
\end{figure*}

Although the frequencies of the extensional bound states in Figs.~\ref{fig:rod_bound}(c) and~\ref{fig:rod_bound}(d) and the flexural eigenmodes in Figs.~\ref{fig:rod_extended}(c) and~\ref{fig:rod_extended}(d) are rather close, the mode profiles differ significantly in appearance.
Also, the frequencies of the flexural eigenmodes depend on the specific boundary conditions chosen and other physical parameters, e.g., the total length of the rod.
When the rod length increases, we expect flexural waves to form a near continuum in the frequency spectrum, whereas the frequencies of the extensional bound states would continue to be determined by the quantization condition, Eq.~\eqref{eq:quantization}.
Setting $\lambda_{-} = 0$ and putting $k = 0$ in Eq.~\eqref{eq:rod_ham}, we also see that flexural waves in a rod are gapless and start propagating well below the first bound extensional state.
The zero mode with $\omega = 0$ and $(\zeta, u) \sim (0, 1)$ is a symmetry-protected mode of the rod equations, Eq.~\eqref{eq:rod}, that exists irrespective of the form of the curvature profile $m(x)$.
This mode corresponds to uniform translations along the $x$ (arclength) direction, but it is a physically valid solution only for free boundary conditions of the rod equations.
Given that the flexural waves are gapless, in very long rods, the extensional bound states appear as quasi-bound states in a continuum of flexural waves.

While the extensional bound states seen in our examples do not interact with flexural waves, subtleties may arise in certain cases.
  For instance, it is not impossible to construct curvature profiles such that the bound extensional states can exchange energy with the flexural waves and undergo mode conversion or cause extensional waves to tunnel through regions of high curvature to reach regions of low curvature~\cite{kernes2021,mannattil2023a}.
In both situations, the extensional states gradually spread outward and appear unbound.
That said, the practical relevance of such pathological curvature profiles remains to be investigated.

\section{Extensional limit of the rod equations}
\label{sec:extensional_limit}

In the previous section, using the semiclassical approximation, we showed that extensional waves form bound states in a curved rod.
We could also compute the frequencies of the bound states rather well.
However, finding the shapes of the bound modes using semiclassical asymptotics is a formidable task~\cite{berry1972}.
Instead, in this section, we will use the extensional limit of the rod equations to find the bound states.

We define the extensional limit of the rod equations as the limit in which the average bending energy of a deformed rod is considerably smaller than its stretching energy.
The bending energy in the rod model described in Sec.~\ref{sec:rods} is $B[\partial_{x}^{2} \zeta +m(x)\partial_{x}u(x)]^{2}$~\cite{kernes2021}, which gives rise to the third- and fourth-order spatial derivatives in the rod equations, Eq.~\eqref{eq:rod}.
Upon neglecting this term, we find the extensional limit of the rod equations, which,
after Fourier transforming in time,
take the form
\begin{subequations}
  \begin{align}
  \label{app:eq:rod_ext1}
  m(x)\left[m(x)\zeta(x) - \partial_{x}u(x)\right] &= \omega^{2}\zeta(x),\\
  \partial_{x}\left[m(x)\zeta(x) - \partial_{x}u(x)\right] &= \omega^{2}u(x).
  \label{app:eq:rod_ext2}
  \end{align}
\end{subequations}
The above equations possess a soft-mode solution with $\omega = 0$ satisfying $m(x)\zeta(x) = \partial_{x}u(x)$, which corresponds to all linear isometries that do not stretch the rod to the lowest order.

We now look for bound-state solutions of Eqs.~\eqref{app:eq:rod_ext1} and \eqref{app:eq:rod_ext2} satisfying $\zeta(\pm\infty) = u(\pm\infty) = 0$ with vanishing derivatives at $x = \pm\infty$.
Let us additionally assume that the tangential component $u(x) = \partial_{x}\phi(x)$, where $\phi(x)$ is an unknown differentiable function satisfying%
\footnote{Assuming $\phi(\pm\infty) = 0$ does not result in any loss in generality.
Without this assumption, on integrating Eq.~\eqref{app:eq:rod_ext2} we see that $\phi(\pm\infty) = C$ (constant).
We then obtain Eq.~\eqref{app:eq:schrodinger} in terms of $\widetilde{\phi}(x) = \phi(x) - C$ with $\zeta = m\widetilde{\phi}$, so that we can work in terms of $\widetilde{\phi}$ alone, which is equivalent to setting $C = 0$.}
$\phi(\pm\infty) = 0$.
Equation~\eqref{app:eq:rod_ext2} then becomes a total derivative, which on integration yields
\begin{equation}
  m(x)\zeta(x) = \partial_{x}^{2}\phi(x) + \omega^{2}\phi(x).
\end{equation}
Putting the above equation in Eq.~\eqref{app:eq:rod_ext1} and ignoring the soft-mode solution with $\omega = 0$, we see that $\zeta(x) = m(x)\phi(x)$, from which we deduce that $\phi(x)$ satisfies a Schr\"{o}dinger-like equation with the potential $m^{2}(x)$, and given by
\begin{equation}
  -\partial_{x}^{2}\phi(x) + m^{2}(x)\phi(x) = \omega^{2}\phi(x).
  \label{app:eq:schrodinger}
\end{equation}
As is well known from elementary quantum mechanics~\cite{buell1995}, Eq.~\eqref{app:eq:schrodinger} always admits a bound-state solution provided that potential $m^{2}(x)$ has a minimum.

For the tanh-type curvature profile with $m = m_{1}(x) = b\tanh(\epsilon x)$, Eq.~\eqref{app:eq:schrodinger} becomes the time-independent Schr\"{o}dinger equation for a particle in a P\"{o}schl--Teller potential~\cite{poschl1933,rosen1932}, whose solutions $\phi(x)$ can be written in terms of the associated Legendre polynomials $P^{\mu}_{\nu}(\,\cdot\,)$~\cite{olver2010}.
On defining
\begin{equation}
  \begin{aligned}
    \kappa(x) &= \tanh(\epsilon x),\\
    \nu &= \tfrac{1}{2}\left[\sqrt{1 + 4b^{2}/\epsilon^{2}} - 1\right],\\
    \mu &= n - \nu \leq 0
  \quad(n \in \mathbb{N}_{0}),\\
  \end{aligned}
  \label{app:eq:poschl_params}
\end{equation}
and solving for $\phi(x)$, we find the (unnormalized) components $\zeta(x) = m(x)\phi(x)$, $u(x) = \partial_{x}\phi(x)$, and the quantized frequencies to be
\begin{equation}
  \begin{aligned}
    \zeta(x) &= b\kappa(x)P^{\mu}_{\nu}\left[\kappa(x)\right],\\
    u(x) &= \partial_{x}P^{\mu}_{\nu}\left[\kappa(x)\right],\\
    \omega^{2} &= b^{2} - \epsilon^{2}\mu^{2}.
  \end{aligned}
  \label{app:eq:poschl}
\end{equation}
Figure~\ref{fig:poschl} presents a comparison of the bound states described by Eq.~\eqref{app:eq:poschl} and the numerical results for extensional bound states.
The agreement is remarkable given the fact that the numerical results were obtained by solving the full wave equations, Eq.~\eqref{eq:rod}, without employing any additional approximations.
From Eq.~\eqref{app:eq:poschl_params} we also see that maximum number of bound states is $n_{\text{max}} \approx \nu = b/\epsilon + \mathcal{O}(b^{2}/\epsilon^{2})$, which agrees with the semiclassical prediction in Eq.~\eqref{eq:rod_maxb}.

We are currently unaware of an exact solution of Eq.~\eqref{app:eq:schrodinger} for the potential $\smash{m^{2}_{2}(x) = \left[b - (b - a)\sech(\epsilon x)\right]^{2}}$ corresponding to a sech-type curvature profile, even though exact solutions for similar potentials exist~\cite{lemieux1969,nieto1978,ishkhanyan2018}.
Although we have had success in approximating this potential as a modified P\"{o}schl--Teller potential~\cite{mannattil2023a},
as the results are qualitatively similar, we do not report them here.

At this juncture, let us remark that Eq.~\eqref{app:eq:schrodinger} can also be derived by diagonalizing the full wave equation in Eq.~\eqref{eq:rod} using the asymptotic method outlined by Littlejohn and coworkers~\cite{littlejohn1991,littlejohn1991a,weigert1993}, without appealing to any physical arguments.
In this method, the operators of the diagonalized equations in symbol form are the two eigenvalues $\lambda_{\pm}$ of the dispersion matrix $\mathsf{D}^{(0)}$.
To find the extensional bound states, we focus on $\lambda_{+}$, the ray Hamiltonian for extensional waves.
Instead of using $\lambda_{+}$ from Eq.~\eqref{eq:rod_ham}, which involves a radical expression, we use the approximate version $\lambda_{+} \approx k^{2} - m^{2} - \omega^{2}$ [see Eqs.~\eqref{eq:rod_kgm} and \eqref{eq:rod_klm}].
Next, we promote $k \to \hat{k} = -i\partial_{x}$, which ``quantizes'' the ray Hamiltonian $\lambda_{+}$ and we obtain Eq.~\eqref{app:eq:schrodinger} for an unknown wave function $\phi(x)$.
The full wave field can be recovered from $\phi(x)$ using $(\zeta, u) \sim \hat{\tau}_{+}\phi(x)$, where $\hat{\tau}_{+}$ is a two-component operator obtained by promoting $k \to \hat{k}$ in the symbol form of the polarization vector $\tau_{+}$.
From Eq.~\eqref{eq:rod_tau}, we see that $\tau_{+} \approx [m(x),\, ik]$, so $\hat{\tau}_{+} \approx [m(x),\, \partial_{x}]$.
We then find $\zeta(x) \sim m(x)\phi(x)$ and $u(x) \sim \partial_{x}\phi(x)$, consistent with our analysis.

\section{Wave localization in a curved shell}
\label{sec:shell}

We now turn to the localization of waves in singly curved shells.
One of the simplest shell theories that involve a curvature-mediated coupling between the tangential and normal components of the displacement field is the widely used Donnell--Yu shell model~\cite{donnell1933,yu1955}.
It is simple in the sense that it describes the undulations of a three-dimensional shell solely in terms of the deformation of the shell's two-dimensional midsurface, ignoring higher-order effects and retaining only the lowest-order derivatives.
For an arbitrarily parameterized midsurface, it is easiest to extract the equations of motion from the covariant form of the Donnell--Yu shell equations derived by \citet{pierce1993a,pierce1993}.

\subsection{Equations of motion}

We consider the middle surface of the shell to be a generalized cylinder~\cite{pressley2010} obtained by translating a plane curve $\bm{\sigma}: \mathcal{X} \to \mathbb{R}^{3}$, parameterized by $x \in \mathcal{X} \subset \mathbb{R}$, perpendicular to the plane containing it.
Thus, the shell is defined by $\bm{\Sigma}: \mathcal{X}\times\mathcal{Y} \to \mathbb{R}^{3}$, $\bm{\Sigma}(x, y) = \bm{\sigma}(x) + y{\bm{e}}_{y}$, where $\bm{e}_{y} = \partial_{y}\bm{\Sigma}$ is a constant unit vector perpendicular to the plane containing $\bm{\sigma}(x)$.
Also, $y \in \mathcal{Y}$ is the coordinate along $\bm{e}_{y}$.
For simplicity, and to make comparisons with the rod equations easier, we assume that $x$ is the arclength.
So, $\bm{e}_{x} = \partial_{x}\bm{\Sigma} = \bm{t}$ is the unit tangent along the curve $\bm{\sigma}$.
Furthermore, we orient $\bm{e}_{y}$ such that surface normal $\bm{e}_{x}\times\bm{e}_{y}$ coincides with the normal $\bm{n}$ to $\bm{\sigma}$.
Then, the only nonvanishing principal curvature of the shell is equal to the curve's signed curvature $m(x)$.
Propagating waves displace the shell from $\bm{\Sigma} \to \bm{\Sigma} + \delta\bm{\Sigma}$.
We write the displacement field $\delta\bm{\Sigma}$ as $\delta\bm{\Sigma} = u\bm{e}_{x} + v\bm{e}_{y} + \zeta\bm{n}$.
Assuming no external forces, the dynamic Donnell--Yu equations are
\begin{widetext}
\vskip-\baselineskip
\begin{subequations}
  \begin{align}
    \varrho \frac{\partial^{2}\zeta}{\partial t^{2}} &= -\widetilde{B}\Delta^{2}\zeta - \widetilde{E}m^{2}(x) \zeta + \widetilde{E}m(x)\left(\frac{\partial u}{\partial x} + \eta\frac{\partial v}{\partial y}\right),\\
    \frac{\varrho}{\widetilde{E}} \frac{\partial^{2}u}{\partial t^{2}} &= -\frac{\partial\left[m(x)\zeta\right]}{\partial x} + \frac{\partial^{2}u}{\partial x^2} + \frac{(1-\eta)}{2}\frac{\partial^{2}u}{\partial y^{2}} + \frac{(1+\eta)}{2}\frac{\partial^{2}v}{\partial x \partial y},\\
    \frac{\varrho}{\widetilde{E}} \frac{\partial^{2}v}{\partial t^{2}}&= -\eta m(x)\frac{\partial \zeta}{\partial y} + \frac{(1+\eta)}{2}\frac{\partial^{2}u}{\partial x \partial y} + \frac{(1-\eta)}{2}\frac{\partial^{2}v}{\partial x^{2}} + \frac{\partial^{2}v}{\partial y^2}.
  \end{align}
\end{subequations}
Here $\Delta$ is the two-dimensional Laplacian, $\eta$ is the Poisson's ratio and $\varrho$ is the density per unit area.
Also, the extensional stiffness $\widetilde{K} = Yh/(1-\eta^{2})$ and bending stiffness $\widetilde{B} = Yh^{3}/\left[12(1-\eta^{2})\right]$, with $Y$ being the Young's modulus and $h$ being the thickness of the shell.
On setting the length units to $\sqrt{\smash[b]{\widetilde{B}/\widetilde{K}}}$ and time units to $\sqrt{\smash[b]{\widetilde{B}\varrho/\widetilde{K}}}$, we arrive at the nondimensional form of these equations, which in matrix form reads
\begin{equation}
  \partial^{2}_{t}
  \begin{pmatrix}
    \zeta\\
    u\\
    v
  \end{pmatrix}
  +
  \widehat{\mathsf{H}}
  \begin{pmatrix}
    \zeta\\
    u\\
    v
  \end{pmatrix} = 0,\enspace
  \text{where}\enspace
  \widehat{\mathsf{H}} =
  \begin{pmatrix}
    \Delta^{2} + m^{2}(x) & -m(x)\partial_x & -\eta m(x) \partial_{y}\\
    m(x)\partial_{x} + m'(x) & -\partial_{x}^{2} - \tfrac{1}{2}(1-\eta)\partial^{2}_{y} & -\tfrac{1}{2}(1+\eta)\partial_{x}\partial_{y}\\
    \eta m(x)\partial_{y} & -\tfrac{1}{2}(1+\eta)\partial_{x}\partial_{y} & -\tfrac{1}{2}(1-\eta)\partial_{x}^{2} - \partial_{y}^{2}
  \end{pmatrix}.
  \label{eq:shell_wave_eq}
\end{equation}
A set of equations analogous to the rod equations (see Appendix~\ref{app:rod_simple}, for instance) can be obtained by suppressing the $y$ derivatives in the submatrix obtained by deleting the third row and column of the operator $\widehat{\mathsf{H}}$ given above.

Translation invariance of $\widehat{\mathsf{H}}$ along $y$ lets us look for time-harmonic solutions with the common factor $e^{i(ly - \omega t)}$, where $l$ is the transverse wavenumber in the $y$ direction and $\omega$ is the frequency of oscillation.
This makes the wave field depend only on the coordinate $x$, and makes the transverse wavenumber $l$ an additional parameter of the operator $\widehat{\mathsf{H}}$.
But the operator $\widehat{\mathsf{H}}$ now has complex coefficients, and the components of its eigenmodes are complex functions.
For this reason, while discussing the numerical results, we use a phase convention such that $\zeta, u$ are always real and $v$ imaginary.
Note that such a phase convention will no longer be necessary if we consider shells that are finite in the transverse $y$ direction.

\begin{figure*}
  \begin{center}
    \includegraphics{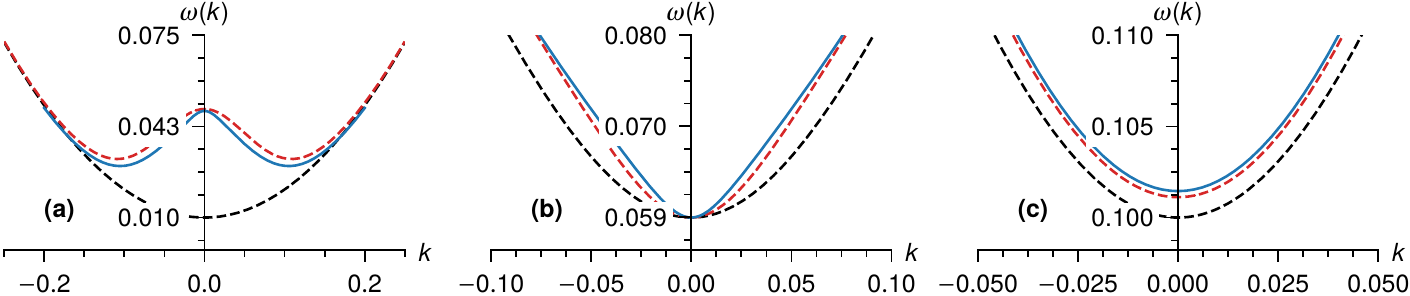}
  \end{center}
  \caption{%
    Dispersion curves for plain waves propagating on a cylinder of constant curvature $m = 0.05$ for (a) flexural, (b) shear, and (c) extensional waves.
    In each plot, the solid blue curves represent the actual dispersion curves obtained by finding $\omega$ from Eq.~\eqref{eq:shell_det}.
    The dashed red curves represent the approximate dispersion relation in Eq.~\eqref{eq:shell_disp_approx}.
    Also, the lower dashed curves, depicted in black, indicate the dispersion curves for an uncurved flat plate, i.e., when $m = 0$ [Eq.~\eqref{eq:shell_disp_zero}].
    The transverse wavenumber $l = 0.1$ and Poisson's ratio $\eta = 0.3$ for all curves.
  }
  \label{fig:shell_disp}
\end{figure*}

Similar to what we did for the rod equations, we perform a change of variables $x \to \epsilon^{-1}x$ and recast the spatial derivatives in terms of the momentum operator $\hat{k}$.
Finally, we find the dispersion matrix
\begin{equation}
\mathsf{D}^{(0)} =
\begin{pmatrix}
  (k^{2} + l^{2})^{2} + m^{2}(x) - \omega^{2} & -i k m(x) & -i\eta l m(x)\\
  ik m(x) & k^{2} + \tfrac{1}{2}(1-\eta)l^{2} - \omega^{2} & \tfrac{1}{2}(1+\eta)kl\\
  i\eta l m(x) & \tfrac{1}{2}(1 + \eta)kl & \tfrac{1}{2}(1-\eta)k^{2} + l^{2} - \omega^{2}
\end{pmatrix}
\end{equation}
and its first-order correction
\begin{equation}
\mathsf{D}^{(1)} =
\tfrac{1}{2}
\begin{pmatrix}
  0 & m'(x) & 0\\
  m'(x) & 0 & 0\\
  0 & 0 & 0
\end{pmatrix}.
\label{eq:shell_disp_matrices}
\end{equation}
For later analysis, it is also useful to note down the determinant of $\mathsf{D}^{(0)}$, which is
\begin{equation}
  \begin{split}
    \det\mathsf{D}^{(0)} &=
      m^{2}(x)\left\{\left[\omega^{2} - \tfrac{1}{2}(1-\eta)l^{2}\right]\left[\omega^{2}-
          (1-\eta^{2})l^{2}\right] -\tfrac{1}{2}(1-\eta)k^{2}\omega^{2}\right\}\\
                         &\quad-\left[\omega^{2} - (k^{2} + l^{2})^{2}\right]
            \left[\omega^{2} - \tfrac{1}{2}(1-\eta)(k^{2} + l^{2})\right]\left[\omega^{2} - (k^{2} + l^{2})\right].
  \end{split}
  \label{eq:shell_det}
\end{equation}
\end{widetext}
Before we continue, it is insightful to examine the dispersion relations for plain waves propagating on singly curved shells of constant curvature.

\subsection{Shells of constant curvature}

\addlines

\begin{figure*}
  \begin{center}
    \includegraphics{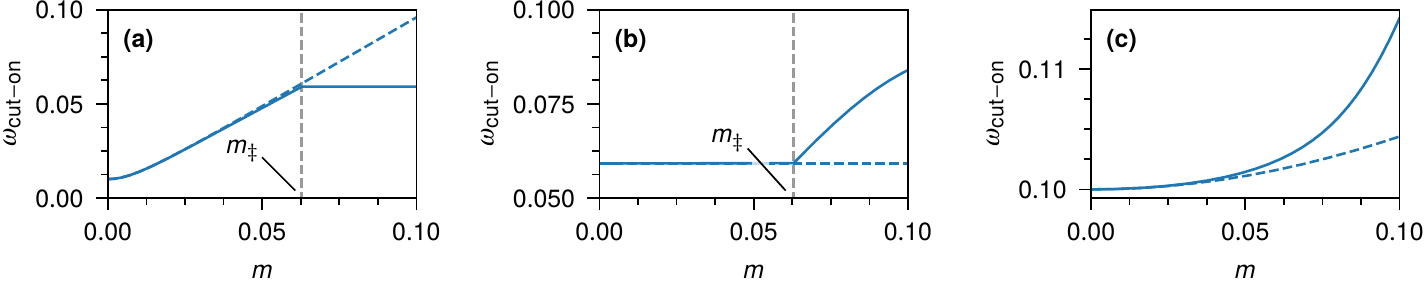}
  \end{center}
  \caption{%
    Cut-on frequencies for a singly curved shell as a function of the curvature $m$ for (a) flexural, (b) shear, and (c) extensional waves.
    The blue dashed curves represent the cut-on frequency predicted by Eq.~\eqref{eq:shell_disp_approx}, which holds only when $m$ is small.
    The solid curves represent the actual cut-on frequency obtained by finding $\omega$ from Eq.~\eqref{eq:shell_det} using the general cubic formula, and taking the limit $k \to 0$ numerically.
    The vertical gray guideline represents $m = m_{\ddagger}$.
    The transverse wavenumber $l = 0.1$ and Poisson's ratio $\eta = 0.3$ in all plots.
  }
  \label{fig:shell_gap}
\end{figure*}

First we analyze the zero curvature limit, i.e., when the shell becomes a flat plate.
In this limit, the wave equation, Eq.~\eqref{eq:shell_wave_eq}, decouples into two equations, the first of which involves only the normal component $\zeta$, and represents flexural waves.
The second equation, representing extensional and shear waves, only involves the tangential components $u$ and $v$.
Shear waves propagate transversely to the in-plane wave vector $k\bm{e}_{x} + l\bm{e}_{y}$, whereas extensional waves are longitudinal to it.
Setting $m = 0$ in Eq.~\eqref{eq:shell_det}, we find the following flat-plate dispersion relations
\begin{equation}
  \begin{aligned}
    \omega_{0}^{2} &= \left(k^{2} + l^{2}\right)^{2},\\
    \omega_{0}^{2} &= \tfrac{1}{2}\left(1-\eta\right)\left(k^{2} + l^{2}\right),\\
    \omega_{0}^{2} &= \left(k^{2} + l^{2}\right),
  \end{aligned}
  \label{eq:shell_disp_zero}
\end{equation}
which we recognize as the dispersion relations of flexural, shear, and extensional waves, respectively~\cite{fung1965}.
The above dispersion relations for a fixed transverse wavenumber $l$ are indicated by the dashed black curves in Fig.~\ref{fig:shell_disp}.
Because $l$ is nonzero, there is a gap in all the dispersion curves and a corresponding nonzero cut-on frequency for each of these waves.

For nonzero but constant curvature, plain waves continue to propagate on the shell, which now becomes part of a thin cylinder.
For simplicity, we shall continue to call these waves as being flexural, extensional, or shear in nature.
However, as with the curved rod, we expect some amount of mixing of the tangential and normal displacements due to nonzero curvature.
Also, to find the dispersion relations from Eq.~\eqref{eq:shell_det} we now need to use the general cubic formula~\cite{olver2010}, which results in unwieldy analytical expressions~\cite{mannattil2023a}.
As an example, we find the dispersion curves at a curvature value of $m = 0.05$, which are indicated by the solid lines in Fig.~\ref{fig:shell_disp}.
We make two observations on comparing these curves with the dispersion curves for a flat plate:
(i) the gap in the dispersion curves for flexural and extensional waves [Figs.~\ref{fig:shell_disp}(a) and~\ref{fig:shell_disp}(c)] have increased, and the flexural dispersion curve now has a double-well appearance;
(ii) although the dispersion curve for shear waves has changed in appearance [Fig.~\ref{fig:shell_disp}(b)], the gap remains the same and the cut-on frequency remains unchanged.
The cut-on frequencies at nonzero $m$ can be computed from Eq.~\eqref{eq:shell_det} after setting $k = 0$, and we find three roots:

\begin{equation}
\begin{aligned}
  \omega^{2}_{\text{cut-on}} = \tfrac{1}{2}\biggl[\!&\pm \sqrt{\left(l^{2} - l^{4} - m^{2}\right)^{2} + 4\eta^{2}l^{2}m^{2}}\\
                                                   &+ l^{2} + l^{4} + m^{2}\biggr]
                             \quad\text{and}\quad\tfrac{1}{2}(1-\eta)l^{2}.
\end{aligned}
\label{eq:shell_cuton}
\end{equation}
Our intuition and a series expansion in $m$ suggests that the lowest of the first two roots above must be associated with flexural waves and the highest root must be associated with extensional waves.
The third root, which is independent of the curvature $m$, must then correspond to the cut-on frequency for shear waves.
As we shall see, this association is only correct at very low curvatures.

Although the exact dispersion relations for nonzero curvature are unwieldy, we can find an approximate expression for the dispersion relations in the very weak curvature limit.
To this end, we write the roots $\omega^{2}$ as a regular perturbation series~\cite{bender1978} in even powers of $m$, i.e.,
\begin{equation}
  \omega^{2}(k, l) = \omega^{2}_{0}(k, l) + \sum_{n = 1}^{\infty} m^{2n}Q_{n}(k, l),
  \label{eq:shell_perturb}
\end{equation}
where $\omega_{0}$ is one of the three roots in Eq.~\eqref{eq:shell_disp_zero} and $Q_{n}(k, l)$ are coefficients to the correction terms that we have to determine.
Since the curvature is assumed to be very weak, the dispersion relations obtained this way can be associated with a wave type based on the choice we make for $\omega_{0}$.
Putting Eq.~\eqref{eq:shell_perturb} in Eq.~\eqref{eq:shell_det}, and dropping powers of $k$ and $l$ in comparison to unity, we find, to $\mathcal{O}(m^{4})$,
\begin{equation}
  \omega^{2} \simeq
  \begin{dcases}
    \left(k^{2} + l^{2}\right)^{2} + \left(1 - \eta^{2}\right)m^{2}\frac{l^{4}}{\left(k^{2} + l^{2}\right)^{2}},\\
    \tfrac{1}{2}\left(1-\eta\right)\left(k^{2} + l^{2}\right) + 2\left(1 - \eta\right)m^{2}\frac{k^{2}l^{2}}{\left(k^{2} + l^{2}\right)^{2}},\\
    (k^{2} + l^{2}) + m^{2}\frac{\left(k^{2} + \eta l^{2}\right)^{2}}{\left(k^{2} + l^{2}\right)^{2}},
  \end{dcases}
  \label{eq:shell_disp_approx}
\end{equation}
which we identify as the approximate dispersion relations of flexural, shear, and extensional waves, respectively.
Similar expressions are also found in the literature, where they have been derived using alternative arguments~\cite{germogenova1973,pierce1993,norris1994,rebinsky1996}.
Also, in their analytical characterization of bound waves in a musical saw, \citet{shankar2022} works exclusively with the above approximate dispersion relation for flexural waves, which has also been observed experimentally~\cite{williams1990}.
In Fig.~\ref{fig:shell_disp}, the approximate dispersion relations for $m = 0.05$ are indicated by the red dashed curves, from which we can see that Eq.~\eqref{eq:shell_disp_approx} captures the true dispersion relations to a reasonably good accuracy.
Finally, although we only consider waves with nonzero $l$ in our analysis, from Eq.~\eqref{eq:shell_disp_approx}, we see that when $l=0$, both flexural and shear waves become gapless with the associated zero modes corresponding to uniform translations along $x$ and $y$.

Equation~\eqref{eq:shell_disp_approx} must break down beyond a certain value of the curvature.
Indeed, we only expect it to capture the true dispersion when the $\mathcal{O}(m^{2})$ correction term in Eq.~\eqref{eq:shell_perturb} is smaller than $\omega_{0}^{2}$, and more conservatively, only when $m \ll l^{2} + k^{2}$.
We would expect the dispersion relation to deviate significantly from Eq.~\eqref{eq:shell_disp_approx} as $m$ increases.
In fact, for small $k$ and $l$, Eq.~\eqref{eq:shell_disp_approx} completely breaks down at a curvature at which the cut-on frequencies for flexural and extensional waves become equal.
For nonzero $l$, from Eq.~\eqref{eq:shell_cuton}, we see that this happens at a curvature value
\begin{equation}
  \begin{aligned}
    m_{\ddag}^{2} &= \frac{(1 + \eta)l^{2}\left[\tfrac{1}{2}(1 - \eta) - l^{2}\right]}{\left(1 + 2\eta\right)\left(1 - \eta\right)}\\
                  &= \tfrac{1}{2}\left(\frac{1+\eta}{1 + 2\eta}\right)l^{2} + \mathcal{O}(l^{4}).
  \end{aligned}
  \label{eq:shell_mdag}
\end{equation}
We graphically demonstrate this in Fig.~\ref{fig:shell_gap}, from which we can see that the expressions for the cut-on frequencies for shear and flexural waves get interchanged at large $m$.
Hence, for $m > m_{\ddag}$, the lowest of the first two cut-on frequencies in Eq.~\eqref{eq:shell_cuton}, should be associated with shear waves, whose dispersion curves would now have a curvature-dependent gap.
The cut-on frequency for flexural waves, however, would now be equal to $\tfrac{1}{2}(1-\eta)l^{2}$.
We remark that this switching of the cut-on frequencies does not happen if $l^{2} > \tfrac{1}{2}(1-\eta)$ as Eq.~\eqref{eq:shell_mdag} fails to have a real root.
Also, as the cut-on frequencies of flexural and shear waves are equal for $m = m_{\ddag}$, in principle, mode conversion can occur close to the entire $k$ axis on the phase plane.
Despite such a possibility, we did not observe any discernible effects of mode conversion in our numerical experiments, and we shall it ignore in our analysis.

\begin{figure*}
  \begin{center}
    \includegraphics{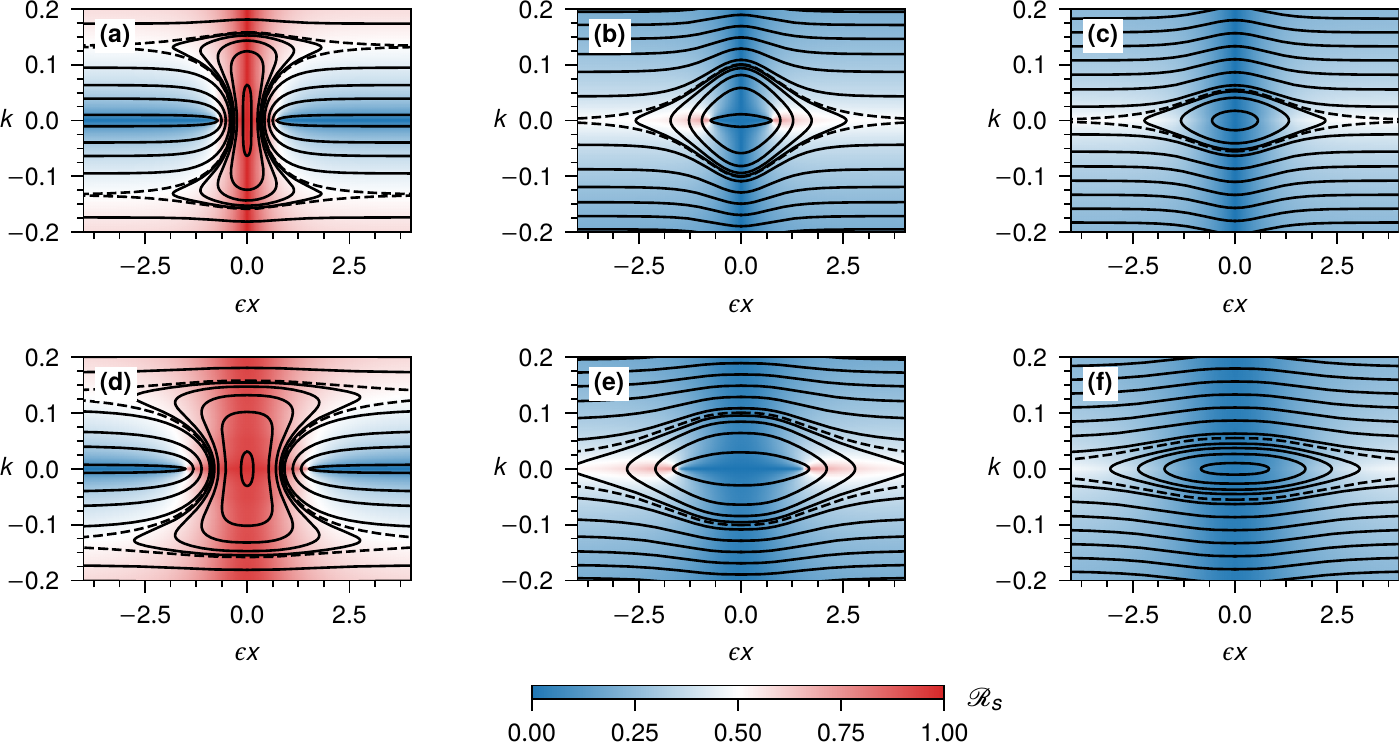}
  \end{center}
  \caption{%
    Ray trajectories for a curved shell with a tanh-type curvature profile (top panels) and a sech-type curvature profile (bottom panels) showcasing (a), (d) flexural waves, (b), (e) shear waves, and (c), (f) extensional waves.
    In all figures the closed curves represent rays associated with quantized bound states and the dashed black curves represent the highest frequency for which there is a bound state.
    The phase portraits have also been color coded with the ratio $\mathscr{R}_{s}$ defined in Eq.~\eqref{eq:shell_ratio}.
  }
  \label{fig:shell_rays}
\end{figure*}

\subsection{Shells with varying curvature}

We shall now consider singly curved shells with varying curvature profiles.
Given the complexity of the general dispersion relations and the myriad of subtleties, we shall, however, perform a less exhaustive analysis compared to what we did for the rod.
We shall only look at a limited number of examples, and to simplify matters, we set the transverse wavenumber $l = 0.1$ and Poisson's ratio $\eta = 0.3$ (corresponding to that of steel) throughout.
As for the rod, we will consider both tanh-type and sech-type curvature profiles, but we assume that the largest absolute curvature $b > m_{\ddag}$.
This would let us examine the problem beyond the range of validity of the approximate dispersion relations in Eq.~\eqref{eq:shell_disp_approx}.
For the sech-type curvature profile, we additionally assume that the smallest curvature $a < m_{\ddag}$.
With $l = 0.1$ and $\eta = 0.3$, we have $m_{\ddag} \approx 0.06$, and these assumptions are satisfied by the choices $a = 0.01$ and $b = 0.1$ that we made for the rod, and we use them for the shell as well.

From our earlier analysis, we saw that the spectral gap in the dispersion relation for all three wave polarizations grows with increasing curvature.
Now, consider a wave traveling from a region of low curvature to one of high curvature.
As the wave moves, at some point, the frequency of the wave would fall below the local cut-on frequency of waves, where it gets reflected back.
Additionally, one can use the approximate dispersion relations in Eq.~\eqref{eq:shell_disp_approx} to show that, analogous to the curved rod, regions of low curvature act as regions of high refractive index, and vice versa~\cite{evans2013}.
Intuitively, we therefore expect bound states to occur for all three wave polarizations.

\begin{figure*}[t]
  \begin{center}
    \includegraphics{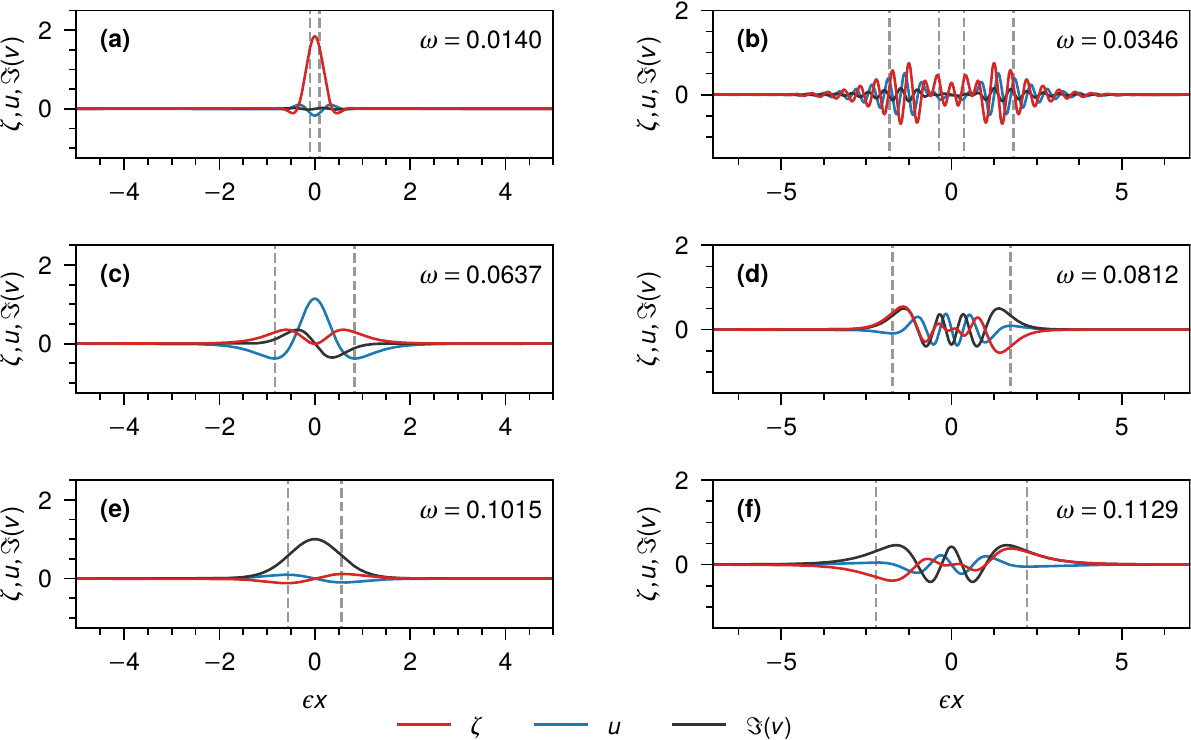}
  \end{center}
  \caption{
    Numerical eigenmodes of a curved shell with a tanh-type curvature profile showing (a),(b) flexural, (c),(d) shear, and (e),(f) extensional bound states.
  In our phase convention, $\zeta$ and $u$ are always real, whereas $v$ is always complex, which is why we only show its imaginary component $\Im(v)$.
  Left panels depict low-frequency bound states, whereas the ones on the right panels have a higher frequency.
  The dashed vertical lines indicate the locations of the caustics.
}
  \label{fig:shell_modes_tanh}
\end{figure*}

For ray analysis, we usually directly work with the eigenvalues of the dispersion matrix $\mathsf{D}^{(0)}$.
This proves to be difficult for the shell as the expressions for the eigenvalues are unwieldy.
In the absence of mode conversion, however, only one eigenvalue, say $\lambda$, will vanish at a given phase-space point $(x, k)$, causing the determinant $\det\mathsf{D}^{(0)}$ to vanish as well.
Hence, the rays determined by $\lambda(x, k; \omega) = 0$ and $\det\mathsf{D}^{(0)} = 0$ are identical, allowing us to use $\det\mathsf{D}^{(0)}$ as the ray Hamiltonian.
Using $\det\mathsf{D}^{(0)}$ instead of $\lambda$ would amount to a trivial reparameterization of Hamilton's equations~\cite{tracy2014}.
Judicious choices of the initial conditions, i.e., the coordinates $(x, k)$ and frequency $\omega$, obtained from the local dispersion curves at a given $(x, k)$, would determine the type of wave the ray represents.

When analyzing a given wave type, it is also useful to compute an amplitude ratio, analogous to the one we used for the curved rod, and defined by
\begin{equation}
  \mathscr{R}_{s} = \frac{\abs{\zeta}}{\abs{\zeta} + \abs{u} + \abs{v}} \sim \frac{\abs{\tau_{1}}}{\abs{\tau_{1}} + \abs{\tau_{2}} + \abs{\tau_{3}}}.
  \label{eq:shell_ratio}
\end{equation}
Above, we have also made use of the fact that the wave field is asymptotic to the polarization vector $\tau$ to write $\mathscr{R}_{s}$ in terms of the components of $\tau$.
With the above definition, flexural waves in a flat plate have $\mathscr{R}_{s} = 1$, whereas both shear and extensional waves have $\mathscr{R}_{s} = 0$.
In a curved shell, because we expect both the normal and tangential components of the wave field to be significant, $\mathscr{R}_{s}$ for the three wave polarizations would deviate from their flat-plate counterparts.
For all three wave types, a significant amount of normal and tangential contribution to the displacement field is indicated by values of $\mathscr{R}_{s}$ in the range $1/3 \lesssim \mathscr{R}_{s} \lesssim 1/2$.
We discuss flexural waves~first.

\subsubsection{Flexural waves}

Our intuitive expectation of flexural bound states is confirmed by the actual ray trajectories for flexural waves showcased in Figs.~\ref{fig:shell_rays}(a) and~\ref{fig:shell_rays}(d).
For frequencies slightly above the cut-on frequency at $x = 0$, the rays appear in the form of closed, vertically elongated orbits that remain confined to a region where the curvature is very small.
At larger frequencies, the rays begin to enter regions of higher curvature, and orbits change from being elliptical to highly eccentric, ``peanut''-shaped curves.
For these orbits, we have a total of six caustics where $\dot{x} = 0$.
[Compare Figs.~\ref{fig:shell_rays}(a) and~\ref{fig:shell_rays}(d) with Fig.~\ref{fig:caustic}(b).]
Two of these caustics, which are on the $x$ axis, are the usual classical turning points where $k = 0$.
At the other four caustics $k \neq 0$, and they arise due to the two double-well minima in the (local) dispersion curves where $\dd{\omega}/\dd{k} = 0$ [see Fig.~\ref{fig:shell_disp}(a), for example].

For the peanut-shaped orbits, bound states do not occur beyond a frequency where the four caustics with $k \neq 0$ get pushed to $x = \pm \infty$.
Since the absolute curvature $\abs{m(\pm \infty)} = b$ for both curvature profiles, the largest frequency for which we see a bound state---represented by the dashed rays in Figs.~\ref{fig:shell_rays}(a) and~\ref{fig:shell_rays}(d)---must be the frequency of the double-well minimum in the dispersion curves for $m = b$.
For small enough $b$, using Eq.~\eqref{eq:shell_disp_approx}, we find this minimum to be $\omega^{2} \approx 2\sqrt{l^{4}b^{2}(1-\eta^{2})}$.
This expression, however, turns out to overestimate the actual minimum for large $b$, and we must find it numerically.
Also, this minimum exists only when $b \gtrsim l^{2}/\sqrt{1-\eta^{2}}$.
For smaller $b$, rays of the bound states remain elliptical in nature, with the classical turning points being the only caustics.

An example profile of a low-frequency flexural bound state is shown in Fig.~\ref{fig:shell_modes_tanh}(a).
This state has negligible tangential components $u$ and $v$, and remains confined to a region of similar extent as the classical turning points.
At higher frequencies, flexural bound states grow beyond the classical turning points and enter regions of higher curvature.
Here curvature effects become more prominent, and the states tend to have both tangential and normal components as seen from Fig.~\ref{fig:shell_modes_tanh}(b), and the color coding of the phase portraits.
They, however, remain confined to a region of similar extent as the four caustics with $\dot{x} = 0$.

It is interesting to note that unlike in the case of rods, flexural waves on shells form bound states.
Intuitively, this may be explained on the basis of the extra transverse degree of freedom possessed by waves on a shell, quantified in our examples by the transverse wave number $l$.
For nonzero $l$, this results in a nonzero spectral gap in the flexural dispersion curves [see Eq.~\eqref{eq:shell_disp_approx} and Fig.~\ref{fig:shell_disp}(a)] leading to the formation of bound flexural states.
Indeed, if we only consider flexural waves that are not oscillatory in the $y$ direction, i.e., when $l = 0$, we recover a situation analogous to that of the rod, and such waves do not form bound states.

\vspace{-\baselineskip}

\subsubsection{Shear waves}

From the phase portraits in Fig.~\ref{fig:shell_rays}(b) and~\ref{fig:shell_rays}(e), we see that shear waves also form bound rays confined between two classical turning points.
As the frequency of the orbits increase, the turning points move to $x = \pm \infty$, where the absolute curvature for both curvature types is $b$.
Thus, the largest frequency for which we observe shear bound states is the shear-wave cut-on frequency for a cylindrical shell of curvature equal to $b$, obtained by setting $m = b$ in the second root of Eq.~\eqref{eq:shell_cuton}.
Beyond this frequency, shear waves form unbound states.
The rays of shear bound states are elongated along the $x$ direction as the spectral gap in the local dispersion curve does not begin increasing until $m(x) > m_{\ddag}$.
For the same reason, shear waves do not get localized if $b < m_{\ddag}$.
It is then natural to wonder if the localization of shear waves seen in the phase portraits is an artifact of having chosen a relatively large value of $b = 0.1$.
But from Eq.~\eqref{eq:shell_mdag} we see that $m_{\ddag}$ can be made arbitrarily small by adjusting the value of transverse wavenumber $l$, so even for small $b$ we would expect shear bound states.

Example profiles of two shear bound states are shown in Figs.~\ref{fig:shell_modes_tanh}(c) and~\ref{fig:shell_modes_tanh}(d).
The first one has a frequency that is only slightly above the shear-wave cut-on frequency at $x=0$ and hence its wavenumber $k$ along the $x$ direction is small compared to its wavenumber $l = 0.1$ along $y$.
Therefore, its (local) wave vector is predominantly in the $y$ direction [see Fig.~\ref{fig:waves}(b)].
Furthermore, as expected from the transverse nature of shear waves, the dominant tangential component is $u$, which is the displacement along $x$.
The second shear bound state shown in Fig.~\ref{fig:shell_modes_tanh}(f) has a higher frequency, causing it to spread to regions of higher curvature, where curvature effects become more prominent.
This can also be inferred from the color coding of Figs.~\ref{fig:shell_rays}(b) and~\ref{fig:shell_rays}(e), which shows that shear bound states develop a significant normal component at higher frequencies.

\subsubsection{Extensional waves}

Color-coded phase portraits of extensional waves indicating bound states are shown in Figs.~\ref{fig:shell_rays}(c) and~\ref{fig:shell_rays}(f).
The caustics for these bound states are the usual classical turning points with $k=0$.
For higher-frequency bound states, these points move to $\pm \infty$, where the absolute curvature is $b$.
Hence, the largest frequency for which we observe extensional bound states must be the cut-on frequency for extensional waves in a shell having a curvature equal to $b$, obtained by putting $m = b$ in the first root in Eq.~\eqref{eq:shell_cuton}.

Example profiles of two extensional bound states are shown in Figs.~\ref{fig:shell_modes_tanh}(e) and~\ref{fig:shell_modes_tanh}(f).
Low-frequency extensional bound states, such as the one in Fig.~\ref{fig:shell_modes_tanh}(e), are expected to displace the shell predominantly in the $y$ direction as seen from the comparatively large values of the tangential component $v$.
The bound state in Fig.~\ref{fig:shell_modes_tanh}(f) has a slightly higher frequency, causing it to spread to regions of higher curvature, where it develops a significant normal component, which we also infer from the color coding of Figs.~\ref{fig:shell_rays}(c) and~\ref{fig:shell_rays}(f).

\subsection{Bound states and quantization}

\begin{figure}
  \begin{center}
    \includegraphics{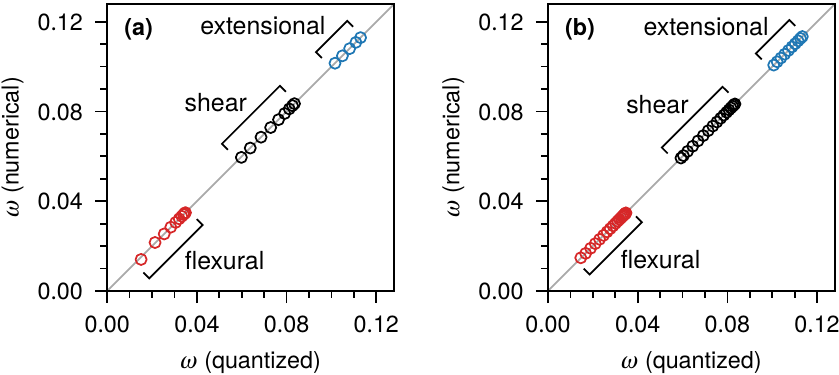}
  \end{center}
  \caption{
    Bound-state frequencies of a curved shell obtained from numerics compared to that obtained through quantization for (a)
    tanh-type curvature profile $m_{1}(x)$ and (b) sech-type curvature profile $m_{2}(x)$.
    For both plots, the gray guideline in the background represents $\omega$ (quantized) = $\omega$ (numerical).
  }
  \label{fig:shell_wkb}
\end{figure}

To find the bound-state frequencies, we first set $k=0$ in Eq.~\eqref{eq:shell_det} and rearrange terms to find that that classical turning points $x^{\star}$ are given by the solutions to the implicit~equation
\begin{equation}
  m^{2}(x^{\star}) = \left[\frac{\omega^{2} - l^{2}}{\omega^{2} - \left(1-\eta^{2}\right)l^{2}}\right]\left(\omega^{2}-l^{4}\right).
  \label{eq:shell_caustic}
\end{equation}
Depending on the value of $\omega$, the turning points found using the above equation could correspond to turning points on the bound rays of all three waves.
We use the same quantization procedure as for the rod to determine the bound-state frequencies (Appendix~\ref{app:numerical}).
Furthermore, the extra phases $\gamma_{\text{G}}$ and $\gamma_{\text{NG}}$ in the quantization condition in Eq.~\eqref{eq:quantization} vanish for the shell equations as well (Appendix~\ref{app:additional_phase}).
The Keller--Maslov index continues to be $\alpha = 2$ for all orbits---including the peanut-shaped orbits with six caustics---as they can be smoothly deformed into a circle centered around the origin~\cite{percival1977}.
From Fig.~\ref{fig:shell_wkb}, we see that the bound-state frequencies obtained through quantization agree rather well with the numerical values for both curvature profiles.

Although waves of all three types form bound states, from our preceding analyses and Fig.~\ref{fig:shell_wkb}, we see that flexural bound states appear first, followed by shear and extensional bound states.
Thus, in very long shells, shear waves form bound states that lie in a quasi-continuum of flexural waves spread across the shell.
Likewise, extensional bound states would lie in a quasi-continuum of unbound flexural and shear waves.
Similar to the curved rod, the bound states of the curved shell are also of definite parity when the curvature profile $m(x)$ is odd or even.
More specifically, when $m(x)$ is even, the components $\zeta(x)$ and $v(x)$ have the same parity, with $u(x)$ having the opposite parity.
For odd $m(x)$, however, $\zeta(x)$ and $u(x)$ have the same parity, with $v(x)$ having the opposite parity, as can be seen from the example bound states in Fig.~\ref{fig:shell_modes_tanh}.

The existence of bound states in shells around points where the absolute curvature has a minimum complements the prediction by \citet{mohammed2021} regarding the localization of flexural waves in similar shells around points of maximal curvature.
Given the qualitative similarity between the bound states observed with such a profile and those seen in this paper, they have been discussed elsewhere~\cite{mannattil2023a}.

Similar to the curved rods we analyzed previously, the two ends of the shells we consider spiral around a cylinder with curvature equal to $b = 0.1$.
It can be easily seen that the ring frequency of extensional waves on such a cylinder is equal to $b$.
Beyond the ring frequency lie the other elementary vibrational modes of the cylinder and from the results in Fig.~\ref{fig:shell_wkb}, we see that almost all the bound states of the shell lie below these elementary modes.
But note that none of these elementary modes displace the shell in the transverse direction.
In comparison, since $l$ is nonzero for the bound modes, these modes displace the shell in the transverse direction.
Also, some of the higher bound modes have frequencies that lie in between that of these elementary modes.

\vspace{-0.7\baselineskip}

\section{Concluding remarks}
\label{sec:conclusion}

\vspace{-0.7\baselineskip}

\addlines[2]

In this paper we have studied the localization of waves in thin elastic structures induced by variations in the structure's curvature profile.
Wave localization is intuitively expected as it is known that curvature acts as an effective refractive index for such waves~\cite{norris1994,evans2013}.
For both the example structures we considered, bound states develop around points where the structure's absolute curvature has a minimum.
In the case of the shell, flexural, shear, and extensional waves form bound states.
In contrast to shells, bound states in a curved rod (which are always extensional in nature) only exist around points where the absolute curvature has a minimum.
Additionally, the extensional bound states in both structures appear as bound states in a near-continuum of flexural waves.
These findings set the stage for the design of simple devices capable of inducing wave localization without relying on metamaterials with nontrivial microstructure.

Semiclassical approximation and other phase space based methods have continued to provide new insights into a wide range of problems in elastodynamics~\cite{mohammed2022,jose2022,jose2023}.
However, such methods have inherent challenges of their own when used to study multicomponent waves, particularly due to the presence of nontrivial phases in the semiclassical quantization rule.
The rod and shell equations we use in this paper, however, have properties that cause these phases to vanish, making the quantization results remarkably accurate.
Nevertheless, topologically protected waves in continuous media can arise when this phase is nonzero, especially when time-reversal symmetry is broken~\cite{venaille2023}.
For this reason, it is worthwhile to explore the use of semiclassical methods in problems with broken time-reversal symmetry such as those in rotating elastic media~\cite{marijanovic2022}, fluids with odd viscosity~\cite{souslov2019}, and magnetoelastic waves~\cite{banos1956}, where one would generically expect this phase to be nonzero.

\vspace{-\baselineskip}

\begin{acknowledgements}
\vspace{-0.5\baselineskip}
  M.M.~thanks Suraj Shankar for useful conversations.
  The authors also thank the anonymous referees for useful feedback.
  This work has been supported by the NSF Grant No. DMR 2217543.
\end{acknowledgements}

\vspace{-\baselineskip}

\appendix

\section{Additional phases}
\label{app:additional_phase}

In this Appendix, we will look at situations where the extra phase $\gamma$ that appears in the quantization condition in Eq.~\eqref{eq:quantization} vanishes.
As we described in the main text, $\gamma = \gamma_{\text{G}} + \gamma_{\text{NG}}$, where $\gamma_{\text{G}}$ is the term that gives rise to a nonzero geometric phase and $\gamma_{\text{NG}}$ is the other (nongeometric) term.
First, we shall analyze general $N$-component wave equations with an $N\times N$ dispersion matrix $\mathsf{D}^{(0)}$.

\subsection{General wave equations}
\label{app:genwave}

Consider a general $N$-component polarization vector $\tau$ of the dispersion matrix $\mathsf{D}^{(0)}$, given by
\begin{equation}
  \tau =
  \begin{pmatrix}
    r_{1}(x, k) e^{i{\varphi}_{1}(x,k)}\\
    r_{2}(x, k) e^{i{\varphi}_{2}(x, k)}\\
    \vdots\\
    r_{N}(x, k) e^{i{\varphi}_{N}(x, k)}
  \end{pmatrix}
  \label{app:eq:tau}
\end{equation}
Above, we have expressed the $\mu$th component $\tau_{\mu}$ in terms of a real amplitude $r_{\mu}(x, k)$ and a phase $\varphi_{\mu}(x, k)$, both of which are functions of the phase-space coordinates $(x,k)$.
Since $\tau$ is normalized, we have $\Abs{\tau}^{2} = \sum_{\mu=1}^{n} r_{\mu}^{2}(x, k) = 1$ for all $(x, k)$.
Putting Eq.~\eqref{app:eq:tau} in Eq.~\eqref{eq:extra_phases}, we see that the rate of change of the first (geometric) phase $\gamma_{\text{G}}$ is
\begin{equation}
  \begin{aligned}
    \dot{\gamma}_{\text{G}} = i\tau^{*}_{\mu}\left\{\tau_{\mu}, \lambda\right\}
                  &= ir_{\mu}\left\{r_{\mu}, \lambda\right\} - r_{\mu}^{2}\left\{\varphi_{\mu}, \lambda\right\}\\
                 &= (i/2)\left\{\Abs{\tau}^{2}, \lambda\right\} - r_{\mu}^{2}\left\{\varphi_{\mu}, \lambda\right\}\\
                 &= -r_{\mu}^{2}\left\{\varphi_{\mu}, \lambda\right\}.
  \end{aligned}
\end{equation}
In the last step above, we have made use of the fact that $\Abs{\tau} = 1$ always, so the Poisson bracket $\{\Abs{\tau}^{2}, \lambda\}$ vanishes.
Clearly, if the phases $\varphi_{\mu}(x, k)$ are constants, then $\dot{\gamma}_{\text{G}}$ vanishes.
More generally, $\dot{\gamma}_{\text{G}}$ would vanish if all $(x, k)$ dependence in the phases $\varphi_{\mu}$ can be removed by an overall rephasing of $\tau$ (as such a rephasing does not affect the normalization of $\tau$).
In other words, only the relative phases between the components of $\tau$ contribute to $\dot{\gamma}_{\text{G}}$.
From here on we assume that the phases $\varphi_{\mu}$ are constants, so all Poisson brackets involving $\varphi_{\mu}$ can be set to zero.
In that case $\dot{\gamma}_{\text{G}} = 0$ everywhere on the phase space, and the accumulated phase $\gamma_{\text{G}}$ as we move along an orbit can be taken to be zero.

But what about the second (nongeometric) phase $\gamma_{\text{NG}}$?
From Eq.~\eqref{eq:extra_phases} we see that the rate of change of $\gamma_{\text{NG}}$ is given~by
\begin{align}
  \notag\dot{\gamma}_{\text{NG}} &= (i/2) \mathsf{D}^{(0)}_{\mu\nu}\left\{\tau_{\mu}^{*}, \tau_{\nu}\right\} - \tau_{\mu}^{*}\mathsf{D}^{(1)}_{\mu\nu}\tau_{\nu}&&\\
  \notag                   &= (i/2)\sum_{\mu < \nu} \left(\mathsf{D}^{(0)}_{\mu\nu}e^{-i\varphi_{\mu\nu}} - \mathsf{D}_{\mu\nu}^{(0)^{*}}e^{i\varphi_{\mu\nu}}\right)\left\{r_{\mu},r_{\nu}\right\}&&\\
                           &\qquad - \tau_{\mu}^{*}\mathsf{D}^{(1)}_{\mu\nu}\tau_{\nu}.
  \label{app:eq:gamma_NG}
\end{align}
In the last step above, we have used Eq.~\eqref{app:eq:tau} to simplify the first term on the RHS and have defined $\varphi_{\mu\nu} = \varphi_{\mu} - \varphi_{k}$.
We have also made use of the Hermiticity of $\mathsf{D}^{(0)}$ to express the first term in terms of the off-diagonal entries of $\mathsf{D}^{(0)}$.
From Eq.~\eqref{app:eq:gamma_NG} we see that even when the phases $\varphi_{\mu}$ are constants, $\gamma_{\text{NG}}$ could be nonzero.
However, in the following subsection we show that for wave equations of thin elastic structures with certain invariant properties, in addition to a vanishing $\gamma_{\text{G}}$, the phase $\gamma_{\text{NG}}$ vanishes as well.

\subsection{Wave equations of thin elastic structures}

We begin by noting that the rod equations, Eq.~\eqref{eq:rod}, as well as the shell equations, Eq.~\eqref{eq:shell_wave_eq}, remain invariant on simultaneously inverting the sign
of the spatial derivatives and the tangential components of the displacement field, i.e., under $(\partial_{x}, \partial_{y}) \to (-\partial_{x}, -\partial_{y})$ and $(\zeta, u, v) \to (\zeta, -u, -v)$.
This invariance can be traced back to the invariance of the strain expressions
used to derive these equations.
More specifically, it arises when the linearized extensional and bending strains are comprised of terms involving only odd derivatives of $u$ and $v$, and even derivatives of $\zeta$, as in models based on the Kirchhoff--Love assumptions~\cite{shankar2022}.
The same invariance is also found in many higher-order theories of rods~\cite{chidamparam1993,walsh2000} and shells~\cite{doyle2021}.
For these reasons, it is useful to consider a general linear elastodynamic equation involving a three-component wave field $\Psi = (\zeta, u, v)$ and possessing this invariance, and given by
\begin{equation}
  \partial_{t}^{2}\Psi(x, y, t) + \widehat{\mathsf{H}}\Psi(x, y, t) = 0,
\end{equation}
where the wave operator is of the form
\begin{equation}
  \widehat{\mathsf{H}} =
  \begin{pmatrix}
    \widehat{Z} & \widehat{A} & \widehat{B}\\
    \widehat{A}^{\dagger} & \widehat{U} & \widehat{C}\\
    \widehat{B}^{\dagger} & \widehat{C}^{\dagger} & \widehat{V}
  \end{pmatrix}.
  \label{app:eq:wave_eq}
\end{equation}
Above, the entries of $\widehat{\mathsf{H}}$ are linear differential operators comprised of powers of $\partial_{x}$ and $\partial_{y}$.
Also, the diagonal entries $\widehat{Z}$, $\widehat{U}$, and $\widehat{V}$ are Hermitian operators and $\widehat{A}^{\dagger}$ is the Hermitian adjoint of $\widehat{A}$.
Additionally, since Eq.~\eqref{app:eq:wave_eq} represents an elastodynamic system, we assume that the coefficients of all the derivatives in $\widehat{\mathsf{H}}$ are real so that $\Psi$ can be taken to be real as well.

Any invariance possessed by Eq.~\eqref{app:eq:wave_eq} must be shared by the (potential) energy density $\mathscr{J} = \tfrac{1}{2}\Psi\trans\widehat{\mathsf{H}}\Psi$ used to derive it from Hamilton's principle.
For instance, this invariance can be readily seen in the rod energy density in Eq.~\eqref{eq:rod_functional}.
If $\mathscr{J}$ is to be invariant under $(\partial_{x}, \partial_{y}) \to (-\partial_{x}, -\partial_{y})$ and $(\zeta, u, v) \to (\zeta, -u, -v)$ for an arbitrary $\Psi$, the off-diagonal operators $\widehat{A}$ and $\widehat{B}$ must be odd under $(\partial_{x}, \partial_{y}) \to (-\partial_{x}, -\partial_{y})$.
In other words, they can only have terms involving exactly one odd power of $\partial_{x}$ (or $\partial_{y}$).
Odd powers of $\partial_{x},\, \partial_{y}$ acquire complex coefficients when expressed in terms of the momentum operator: $\partial_{x}^{2n+1} = (-1)^{n}i\hat{k}^{2n+1}$.
Meanwhile, coefficients of even powers of $\partial_{x},\, \partial_{y}$ remain real: $\partial_{x}^{2n} = (-1)^{n}\hat{k}^{2n}$.
Using the rules in Eq.~\eqref{eq:weylrules}, we therefore conclude that the lowest-order symbols of the off-diagonal operators  $\widehat{A}$ and $\widehat{B}$ must be purely complex (as $\widehat{\mathsf{H}}$ did not have complex coefficients to begin with).
From Eq.~\eqref{eq:weylrules} we also see that the $\mathcal{O}(\epsilon)$ corrections to these symbols must be real.
Therefore, we can write down the symbols of the operators $\widehat{A}$ and $\widehat{B}$ as
\begin{equation}
  \begin{aligned}
    A &= iA^{(0)} + \epsilon A^{(1)} + \mathcal{O}(\epsilon^{2}),\\
    B &= iB^{(0)} + \epsilon B^{(1)} + \mathcal{O}(\epsilon^{2}).
  \end{aligned}
\end{equation}
where $A^{(0)}, B^{(0)}$, etc., are real functions.
A similar reasoning would reveal that the operator $\widehat{C}$ must be even under $(\partial_{x}, \partial_{y}) \to (-\partial_{x}, -\partial_{y})$, and consequently, its symbol is of the form
\begin{equation}
  C = C^{(0)} + i\epsilon C^{(1)} + \mathcal{O}(\epsilon^{2}),
\end{equation}
where $C^{(0)}$ and $C^{(1)}$ are real.
As $\widehat{\mathsf{H}}$ is a Hermitian operator, the symbols of the diagonal entries are all real and $\mathcal{O}(\epsilon)$ corrections to these symbols must vanish.

For finding the eigenmodes, after Fourier transforming in time, we define $\widehat{\mathsf{D}} = \widehat{\mathsf{H}} - \omega^{2}\mathsf{I}_{3}$ and convert $\widehat{\mathsf{D}}$ to its symbol form $\mathsf{D} = \mathsf{D}^{(0)} + \epsilon\mathsf{D}^{(1)} + \mathcal{O}(\epsilon^{2})$.
From the above discussion, we see that the most general dispersion matrix $\mathsf{D}^{(0)}$ and its $\mathcal{O}(\epsilon)$ correction $\mathsf{D}^{(1)}$ that can be written down are of the form
\begin{equation}
\begin{aligned}
  \mathsf{D}^{(0)} &=
\begin{pmatrix}
  Z^{(0)} - \omega^{2} & i{A}^{(0)} & i{B}^{(0)}\\
  -i{A}^{(0)} & U^{(0)} - \omega^{2} & {C}^{(0)}\\
  -i{B}^{(0)} & C^{(0)} & V^{(0)} - \omega^{2}
\end{pmatrix},\\
  \mathsf{D}^{(1)} &=
\begin{pmatrix}
  0 & {A}^{(1)} & {B}^{(1)}\\
  A^{(1)} & 0 & iC^{(1)} \\
  B^{(1)} & -iC^{(1)} & 0
\end{pmatrix}.
\end{aligned}
\label{app:eq:gen_disp_matrix}
\end{equation}

A polarization vector $\tau$ is defined up to an overall phase and normalization by $\mathsf{D}^{(0)}\tau = 0$.
Direct inspection reveals that for $\mathsf{D}^{(0)}$ defined in Eq.~\eqref{app:eq:gen_disp_matrix}, we can take $\tau$ to be of the form%
\footnote{To see this more explicitly, take $\tau = (r_{1}e^{i\phi_{1}},\, \tau_{2},\, \tau_{3})$ and solve for the components $\tau_{2}$ and $\tau_{3}$ from $\mathsf{D}^{(0)}\tau = 0$.
  Upon rephasing $\tau$ by $e^{-i\phi_{1}}$, we find that $\tau$ is of the general form in Eq.~\eqref{app:eq:tau3}.
}
\begin{equation}
  \tau =
  \begin{pmatrix}
    \phantom{i}\tau_{1}\\
    i\tau_{2}\\
    i\tau_{3}\\
  \end{pmatrix},
  \label{app:eq:tau3}
\end{equation}
where $\tau_{1}$, $\tau_{2}$, and $\tau_{3}$ are \emph{real} functions defined on the phase space.
Clearly, the relative phases $\varphi_{12}$ and $\varphi_{13}$ between the components of $\tau$ are either $\pm \pi/2$ or $0$ (when $\tau_{1}$ or $\tau_{2}$ vanishes).
Likewise, the relative phase $\varphi_{23}$ is either $0$ or $\pi$.
Because the relative phases are constants, from our discussion in the previous subsection, it then follows that the geometric phase $\gamma_{\text{G}} = 0$.
When the matrix $\mathsf{D}^{(1)}$ is of the form in Eq.~\eqref{app:eq:gen_disp_matrix}, using the polarization vector $\tau$ in Eq.~\eqref{app:eq:tau3}, a straightforward computation shows that second term in the expression for $\dot{\gamma}_{\text{NG}}$, Eq.~\eqref{app:eq:gamma_NG}, vanishes.
Next, we note that $\mathsf{D}_{12}^{(0)}e^{-i\varphi_{12}} = \pm A^{(0)}$, $\mathsf{D}_{13}^{(0)}e^{-i\varphi_{13}} = \pm B^{(0)}$, and $\mathsf{D}_{23}^{(0)}e^{-i\varphi_{23}} = \pm C^{(0)}$, are all real.
From Eq.~\eqref{app:eq:gamma_NG} it then follows that $\dot{\gamma}_{\text{NG}}$ vanishes, and we can take $\gamma_{\text{NG}}$ to be zero as well.

The dispersion matrices for the thin shell we considered in Eq.~\eqref{eq:shell_disp_matrices} of the main text is of the form in Eq.~\eqref{app:eq:gen_disp_matrix}, and hence the phase $\gamma = \gamma_{\text{G}} + \gamma_{\text{NG}}$ is zero for the shell.
It can be verified that the dispersion matrices for many higher-order shell theories~\cite{doyle2021} would also be of this form.
For the curved rod, the dispersion matrices in Eq.~\eqref{eq:rod_D} are identical in form to the general dispersion matrices in Eq.~\eqref{app:eq:gen_disp_matrix} once we delete the third row and column.
Proceeding by arguments similar to previous ones, we see that the extra phase $\gamma$ vanishes for the rod as well.

\section{Wave localization in a simpler rod model}
\label{app:rod_simple}

A simpler rod model~\cite{kernes2021,mannattil2023a}, which can be considered as the one-dimensional analog of the Donnell--Yu shell equations can also be used to demonstrate the localization of extensional waves in a curved rod.
This model is derived by dropping the curvature-dependent term $m(x)\partial_{x}u$ in the bending energy in Eq.~\eqref{eq:rod_functional}, but keeping the stretching energy unaltered.
Although this may seem like an excessive approximation, as we outline below, the new equations give nearly the same results as the rod model in Eq.~\eqref{eq:rod}.
In nondimensional form, the simplified rod equations can be written as
\begin{equation}
\partial_{t}^{2}
\begin{pmatrix}
  \zeta\\
  u
\end{pmatrix} +
\widehat{\mathsf{H}}
\begin{pmatrix}
  \zeta\\
  u
\end{pmatrix} = 0,
\label{eq:rod2}
\end{equation}
where
\begin{equation}
\widehat{\mathsf{H}}
=
\begin{pmatrix}
  \partial^{4}_{x} + m^{2}(x) & -m(x)\partial_{x}\\
  m(x)\partial_{x} + m'(x) & -\partial^{2}_{x}
\end{pmatrix}.
\end{equation}
The above equations can also be derived from the Donnell--Yu equations, Eq.~\eqref{eq:shell_wave_eq}, by dimensional reduction.

Following the usual procedure of converting derivatives into operators, and operators into symbols, we find that dispersion matrix for the eigenvalue problem is $\mathsf{D} = \mathsf{D}^{(0)} + \epsilon\mathsf{D}^{(1)}$, where
\begin{equation}
  \begin{aligned}
    \mathsf{D}^{(0)} &=
    \begin{pmatrix}
      {k}^{4} + m^{2}(x) - \omega^{2} & -i m(x){k}\\
      im(x){k} & {k}^{2} - \omega^{2}
    \end{pmatrix},\\
    \mathsf{D}^{(1)} &=
    \tfrac{1}{2}
    \begin{pmatrix}
      0 & m'(x)\\
      m'(x) & 0
    \end{pmatrix}.
  \end{aligned}
  \label{eq:rod2_D}
\end{equation}
The same dispersion matrices as the ones above can be recovered from the dispersion matrices in Eq.~\eqref{eq:rod_D} if we drop $k^{2}$ and $m^{2}$ in comparison to unity.
As we work in the limit were both $k^{2}, m^{2} \ll 1$, this shows why the simpler rod equations are an excellent approximation to the rod model in Eq.~\eqref{eq:rod}.

\begin{figure}
  \begin{center}
    \includegraphics{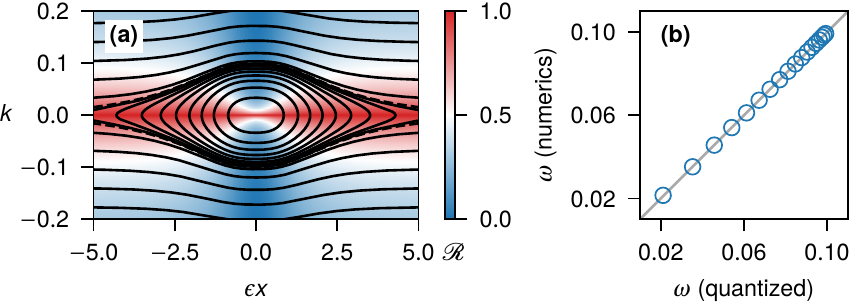}
  \end{center}
  \caption{%
    (a) Ray trajectories for extensional waves described by the simpler rod equations, Eq.~\eqref{eq:rod2}, for the sech-type curvature profile [cf.~Fig.~\ref{fig:rod_bound}(b)].
    (b) Bound-state frequencies obtained from numerics compared to those obtained by quantization.
  }
  \label{fig:rod_simple}
\end{figure}

A straightforward analysis, by inspecting the Hamilton's equations obtained from the eigenvalues of $\mathsf{D}^{(0)}$ in Eq.~\eqref{eq:rod2_D}, reveals again that only extensional waves form bound states with the simpler rod model.
Figure~\ref{fig:rod_simple} shows the ray trajectories for extensional waves described by Eq.~\eqref{eq:rod2}.
Clearly, these rays are nearly identical to those seen previously in Fig.~\ref{fig:rod_bound}, showing again that Eq.~\eqref{eq:rod2} is an excellent approximation to Eq.~\eqref{eq:rod}.
For the same reason, the frequencies of the extensional bound states obtained through quantization and numerics also tend to agree rather well with our previous results; compare, for example, Fig.~\ref{fig:rod_wkb}(b) with Fig.~\ref{fig:rod_simple}(b).
Other aspects of quantization, e.g., the lack of the extra phase $\gamma$, location of the turning points, $m^{2}(x) = \omega^{2}$, etc., also carry over.

\section{Numerical details}
\label{app:numerical}

\subsection{Eigenfrequencies and eigenmodes}

To find the eigenmodes numerically, we solve the rod and shell equations, Eqs.~\eqref{eq:rod} and \eqref{eq:shell_wave_eq}, with $x \in \mathcal{X} = [-1000, 1000]$, using Dedalus~\cite{burns2020} with a Chebyshev spectral decomposition and 2048 modes.%
\footnote{The code we use for all our numerical calculations is publicly available~\cite{github}.}
Bound states are identified by manual examination of the eigenmode profiles.
To test the robustness of the bound states, we independently use clamped, simply supported, and mixed clamped--simply supported boundary conditions for both the rod and the shell.
At the clamped end of a rod, the geometric boundary conditions are $\zeta(x) = \partial_{x}\zeta(x) = u(x) = 0$~\cite{kernes2021}.
For the shell, at the clamped end, we additionally have $v(x) = 0$ as well.
At a simply supported end of a rod, we have the geometric boundary condition $\zeta(x) = u(x) = 0$ and the natural boundary condition $\partial_{x}^{2}\zeta(x) + m(x)u'(x)= 0$ (no bending moment)~\cite{fung1965}.
In the case of a shell, at a simply supported end, we have $\zeta(x) = u(x) = v(x) = 0$ and $\partial_{x}^{2}\zeta(x) - \eta l^{2}\zeta(x) = 0$~\cite{yu1955}.

\subsection{Numerical quantization}
\label{app:quantized}

For computing the quantized frequencies, for a given $n \in \mathbb{N}_{0}$, we start with an approximate guess for the frequency $\omega$ based on the numerical results.
We then numerically integrate the ray equations starting at one of the classical turning points on the $x$ axis, say at $x = -x^{\star}$, until the ray reaches the other turning point at $x = x^{\star}$ [see Fig.~\ref{fig:caustic}(a)].
Next, we compute
\begin{equation}
  n(\omega) = (\pi\epsilon)^{-1} \int_{-x^{\star}}^{x^{\star}} \dd{x}\,k(x) - \tfrac{1}{2}
\end{equation}
using points $\left[x, k(x)\right]$ from the ray trajectory, with the integral evaluated by quadrature.
For a general $\omega$, the estimated $n(\omega)$ will not be integer-valued.
Quantized frequencies $\omega$ can be obtained by solving $n(\omega) = n$ using a numerical root finder.
Alternatively, we could minimize the absolute ``error'' $\abs{n - n(\omega)}$ using random values of $\omega$ spread around the initial guess, and take the quantized frequency to be $\argmin_{\omega}\,\abs{n - n(\omega)}$.
For the results reported in the main text, this error is less than $10^{-10}$.

\bibliography{library,misc}

\end{document}